\documentstyle[12pt]{article}

\newcommand{\bea}{\begin{eqnarray}}
\newcommand{\eea}{\end{eqnarray}}
\newcommand{\ps}{\varphi}

\begin{document}

\setcounter{page}{0} \thispagestyle{empty}

\begin{flushright} {\bf
hep-ph/0208220\\
} \end{flushright}

\vskip 2cm

\begin{center}
{\Large {\bf DGLAP and BFKL equations in the $N=4$ supersymmetric gauge
theory. }} \\[0pt]
\vspace{1.5cm} {\large \ A.V. Kotikov } \\[0pt]
\vspace{0.5cm} {\em Bogoliubov Laboratory of Theoretical Physics \\[0pt]
Joint Institute for Nuclear Research\\[0pt]
141980 Dubna, Russia }\\[0pt]

\vspace{0.5cm} and \\[0pt]
\vspace{0.5cm} {\large \ L.N. Lipatov }\\[0pt]
\vspace{0.5cm} {\em Theoretical Physics Department\\[0pt]
Petersburg Nuclear Physics Institute\\[0pt]
Orlova Roscha, Gatchina\\[0pt]
188300, St. Petersburg, Russia 
}\\[0pt]
\end{center}

\vspace{1.5cm}\noindent

\begin{center}
{\bf Abstract}
\end{center}

We derive the DGLAP and BFKL evolution equations in the $N=4$ supersymmetric
gauge theory in the next-to-leading approximation.
The eigenvalue of
the BFKL kernel in this model turns out to be an analytic function of the
conformal spin $|n|$. Its analytic continuation to negative $|n|$ in the
leading logarithmic approximation 
allows us to obtain residues of
anomalous dimensions $\gamma $ of twist-2 operators in the non-physical
points $j=0,-1,...$ from the BFKL equation in an agreement with their direct
calculation from the DGLAP equation. Moreover, in the multi-color limit of
the $N=4$ model the BFKL and DGLAP dynamics in the leading logarithmic 
approximation 
is integrable for an
arbitrary number of particles. In the next-to-leading approximation
the holomorphic separability of the
Pomeron hamiltonian is violated, but the corresponding Bethe-Salpeter kernel
has the property of a hermitian separability. The main singularities of
anomalous dimensions $\gamma $ at $j=-r$ obtained from the BFKL and DGLAP
equations in the next-to-leading approximation
coincide but our accuracy is not enough to verify an
agreement for residues of subleading poles. %
%
%
%
%
%
%
%
%

{\em PACS:} 12.38.Bx

\newpage


\pagestyle{plain}

\section{Introduction}

\indent
The Balitsky-Fadin-Kuraev-Lipatov (BFKL) equation \cite{BFKL, BL}
and the Dokshitzer-Gribov-Lipatov-Altarelli-Parisi (DGLAP) equation \cite
{DGLAP} are used now for a theoretical description of structure functions of
the deep-inelastic $ep$ scattering (DIS) at small values of the Bjorken
variable $x$. The structure functions are measured by the H1 and ZEUS
collaborations \cite{H1}. The higher order QCD corrections to the splitting
kernels of the DGLAP equation are well known \cite{corAP,MerNeer}.
On the other hand, the calculation of the next-to-leading order (NLO)
corrections to the BFKL kernel was completed comparatively recently \cite
{FL,CaCi,KoLi}.

In supersymmetric gauge theories the structure of the BFKL and DGLAP
equations is significantly simplified. In the case of an extended $N=4$ SUSY
the 
NLO corrections to the BFKL equation were calculated in ref. \cite{KoLi} for
an arbitrary value of the conformal spin $|n|$ in a framework of the
dimensional regularization (DREG) scheme. Below in the section 3 these
results are presented in the dimensional reduction (DRED) scheme \cite{DRED}
which does not violate the supersymmetry. The analyticity of the eigenvalue
of the BFKL kernel as a function of the conformal spin $|n|$ gives a
possibility to relate the DGLAP and BFKL equations in this model, as we show
below (see also \cite{KoLi}). Further, in both schemes (DREG and DRED) this
eigenvalue has the property of the hermitian separability which is similar
to the holomorphic separability (see Section 3). \newline

Let us introduce the unintegrated parton distributions (UnPD) $\varphi
_{a}(x,k_{\bot }^{2})$ (hereafter $a=q,g,\varphi $ for the spinor, vector
and scalar particles, respectively) and the (integrated) parton
distributions (PD) $f_{a}(x,Q^{2})$
\[
f_{a}(x,Q^{2})~=~\int_{k_{\bot }^{2}<Q^{2}}dk_{\bot }^{2}\,\varphi
_{a}(x,k_{\bot }^{2})\,.
\]
In DIS $Q^2=-q^2$ and $x=Q^2/(2pq)$ are the Bjorken variables, $k_{\bot }$
is the transverse component of the parton momentum and $q$ and $p$ are the
virtual photon and hadron momenta, respectively.

The DGLAP equation has a clear probabilistic interpretation and relates the
parton densities with different values of $Q^{2}$:
\begin{eqnarray}
\frac{d}{d\ln {Q^{2}}}f_{a}(x,Q^{2}) &=&-\widetilde{W}_{a}f_{a}(x,Q^{2})+%
\int_{x}^{1}\frac{dy}{y}\sum_{b}\widetilde{W}_{b\rightarrow
a}(x/y)f_{b}(y,Q^{2}),  \nonumber \\
\widetilde{W}_{a} &=&\sum_{b}\int_{0}^{1}dx\,\,x\,\,\widetilde{W}_{a\rightarrow
b}(x),  \label{in2}
\end{eqnarray}
where the second term in the r.h.s. of the equation is the Mellin
convolution of the transition probabilities $\widetilde{W}_{b\rightarrow a}(x)$
and PD $f_{b}(x,Q^{2})$. Usually the first and second terms are unified with
the use of the splitting kernel $W_{b\rightarrow a}$ \cite{DGLAP}:
\begin{eqnarray}
\frac{d}{d\ln {Q^{2}}}f_{a}(x,Q^{2})=\int_{x}^{1}\frac{dy}{y}%
\sum_{b}W_{b\rightarrow a}(x/y)f_{b}(y,Q^{2})\,.  \label{in2.1}
\end{eqnarray}

It is well known, that equation (\ref{in2.1}) is simplified after its Mellin
transformation to the Lorentz spin $j$ representation:
\[
\frac{d}{d\ln {Q^{2}}}f_{a}(j,Q^{2})=\sum_{b}\gamma _{ab}(j)f_{b}(j,Q^{2}),
\]
where
\[
f_{a}(j,Q^{2})=\int_{0}^{1}dx\,\,x^{j-1}f_{a}(x,Q^{2})
\]
are the Mellin momenta of parton distributions. The Mellin moment of the
splitting kernel
\[
\gamma _{ab}(j)=\int_{0}^{1}dx\,\,x^{j-1}W_{b\rightarrow a}(x)
\]
coincides with the anomalous dimension matrix for the twist-2 operators
\footnote{%
As in Ref. \cite{FL,KoLi}, the anomalous dimensions differ from those used
usually in the description of DIS by a factor $(-1/2)$, i.e. $\gamma
_{ab}(j)=(-1/2)\gamma _{ab}^{DIS}(j)$.}. These operators are constructed as
bilinear combinations of the fields which describe corresponding partons
(see Eq. (\ref{in4}) below).

\vskip 0.3cm

On the other hand, the BFKL equation relates the unintegrated gluon
distributions with small values of the Bjorken variable $x$:
\[
\frac{d}{d\ln {(1/x)}}\varphi _{g}(x,k_{\bot }^{2})~=~2\omega (-k_{\bot
}^{2})\,\varphi _{g}(x,k_{\bot }^{2})+\int d^{2}k_{\bot }^{\prime }K(k_{\bot
},k_{\bot }^{\prime })\varphi _{g}(x,k_{\bot }^{2}),
\]
where $\omega (-k_{\bot }^{2})<0$ is the gluon Regge trajectory \cite{BFKL}.
\newline

Let us introduce the local gauge-invariant twist-two operators:
\begin{eqnarray}
O_{\mu _{1},...,\mu _{j}}^{g} &=&\hat{S}G_{\rho \mu _{1}}\,D_{\mu
_{2}}\,D_{\mu _{3}}...\,D_{\mu _{j-1}}G_{\rho \mu _{j}},  \nonumber \\
\widetilde{O}_{\mu _{1},...,\mu _{j}}^{g} &=&\hat{S}G_{\rho \mu _{1}}\,D_{\mu
_{2}}\,D_{\mu _{3}}...\,D_{\mu _{j-1}}\widetilde{G}_{\rho \mu _{j}},  \nonumber
\\
O_{\mu _{1},...,\mu _{j}}^{q} &=&\hat{S}\overline{\Psi }\gamma _{\mu
_{1}}\,D_{\mu _{2}}...\,D_{\mu _{j}}\Psi ,  \nonumber \\
\widetilde{O}_{\mu _{1},...,\mu _{j}}^{q} &=&\hat{S}\overline{\Psi }\gamma
_{5}\,\gamma _{\mu _{1}}\,D_{\mu _{2}}...\,D_{\mu _{j}}\Psi ,  \nonumber \\
O_{\mu _{1},...,\mu _{j}}^{\varphi } &=&\hat{S}\Phi D_{\mu _{1}}\,D_{\mu
_{2}}...\,D_{\mu _{j}}\Phi ,  \label{in4}
\end{eqnarray}
where the spinor $\Psi$ and field tensor $G_{\rho \mu}$ describe gluinos and
gluons, respectively. The last expression is constructed from the covariant
derivatives $D_{\mu}$ of the scalar field $\Phi $ appearing in extended
supersymmetric models. The symbol $\hat{S}$ implies a symmetrization of the
tensor in the Lorentz indices $\mu _{1},...,\mu _{j}$ and a subtraction of
its traces.

The matrix elements of $O_{\mu _{1},...,\mu _{j}}^{a}$ and $\widetilde{O}_{\mu
_{1},...,\mu _{j}}^{a}$ are related to the moments of the distributions $%
f_{a}(x,Q^{2})$ and $\Delta f_{a}(x,Q^{2})$ for unpolarized and polarized
partons in a hadron $h$ in the following way
\begin{eqnarray}
\int_{0}^{1}dx\,x^{j-1}f_{a}(x,Q^{2}) &=&<h|\widetilde{n}^{\mu _{1}}...\,%
\widetilde{n}^{\mu _{j}}\,O_{\mu _{1},...,\mu
_{j}}^{a}|h>\,,~~~~~~~~a=(q,g,\varphi )\,,  \nonumber \\
\int_{0}^{1}dx\,x^{j-1}\Delta f_{a}(x,Q^{2}) &=&<h|\widetilde{n}^{\mu
_{1}}...\,%
\widetilde{n}^{\mu _{j}}\,\widetilde{O}_{\mu _{1},...,\mu
_{j}}^{a}|h>\,,~~~~~~~~a=(q,g)\,.  \label{in5}
\end{eqnarray}
Here the vector $\widetilde{n}^{\mu }$ is light-like: $\widetilde{n}^{2}=0$. In the
deep-inelastic $ep$ scattering we have  $\widetilde{n}^{\mu }\sim q^{\mu
}+xp^{\mu }$.

The 
conformal spin $|n|$ and the quantity $1+\omega $ ($\omega $ is an
eigenvalue of the BFKL kernel) coincide respectively with total numbers of
the transversal and longitudinal Lorentz indices of the tensor $O_{\mu
_{1},...,\mu _{J}}^{a}$ with the rank

\begin{equation}
J=1+\omega +|n|\,.
\end{equation}
Namely, we can introduce the following projectors of this tensor

\begin{eqnarray}
\widetilde n^{\mu_1}...\widetilde n^{\mu_{1+\omega}}\, O_{\mu_1,..., \mu_{1+\omega},
\sigma_1,...,\sigma_{|n|}}^a \,
l^{\sigma_1}_{\bot}...l^{\sigma_{|n|}}_{\bot},  \label{in5.1}
\end{eqnarray}
where the complex transverse vector $l_{\bot}$ is given below
\[
l^{\sigma}_{\bot}~=~ \frac{1}{\sqrt{2}} \left(\delta^{\sigma}_1 +
i\delta^{\sigma}_2 \right),~~~l^2_{\bot}=l_{\bot}p=l_{\bot}q=0 \,.
\]

It is important, that the anomalous dimension matrices $\gamma _{ab}(j)$ and
$\widetilde{\gamma}_{ab}(j)$ for the twist-2 operators $O_{\mu _{1},...,\mu
_{j}}^{a}$ and $\widetilde{O}_{\mu _{1},...,\mu _{j}}^{a}$ do not depend on
various projections of indices due to the Lorentz invariance. But generally
for the mixed projections the Lorentz spin $j$ of the tensor is less than
the total number of its Lorentz indices $J$.

The matrix elements of the light-cone projections are expressed through
solutions $f_a$ of the DGLAP equation (see Eqs.(\ref{in5})). On the other
hand the mixed projections

\begin{eqnarray}
\widetilde{n}^{\mu _{1}}...\,\widetilde{n}^{\mu _{1+\omega }}\,l_{\bot }^{\mu
_{2+\omega }}...l_{\bot }^{\mu _{j}}\,<P|O^{g}_{\mu _{1},...,\mu
_{j}}|P>~\sim ~\int_{0}^{1}dxx^{\omega }\int d^{2}k_{\bot }{\left( \frac{%
k_{\bot }}{|k_{\bot }|}\right) }^{n}\varphi _{g}(x,k_{\bot }^{2}),
\label{in5.2}
\end{eqnarray}
can be found at small $\omega $ in terms of solutions $\varphi _g$ of the
BFKL equation. \newline

Thus, it looks reasonable to extract some additional information concerning
the parton $x$-distributions satisfying the DGLAP equation from the
analogous $k_{\bot }$-distributions satisfying the BFKL equation. Moreover,
due to the fact, that in an extended $N=4$ SUSY the $\beta $-function
vanishes, the $4$-dimensional conformal invariance may allow us to relate
the Regge and Bjorken asymptotics of scattering amplitudes.\newline


Our paper is organized as follows \footnote{%
Some results were presented on the XXXV Winter School (see \cite{KoLi01})}.
In Section 2 we discuss the relation between DGLAP and BFKL equations in the
$N=4$ model obtained by an analytic continuation of the eigenvalue $\omega
^{0}(n,\nu )$ in $|n|$. In Sections 3 and 4 we review shortly our previous
results for the BFKL kernel \cite{KoLi}, rewrite them 
in a framework of the DRED 
scheme and investigate their properties. Section 5 is devoted to an
independent calculation of 
anomalous dimensions in the leading and next-to-leading approximations for
N=4 SUSY with the use of the renormalization group. In Section 6 we discuss
the relation between the DGLAP and BFKL equations in the NLO approximation
for this model. Appendices A, B and C are devoted to the derivation of some
results used below. A summary is given in Conclusion.


\section{Anomalous dimensions of twist-2 operators and their singularities}

\indent
In the leading logarithmic approximation
fermions and scalars do not give any contribution to the BFKL
equation and therefore its integral kernel is the same at all supersymmetric
gauge theories. Due to the M\"{o}bius invariance in the impact parameter
space $\overrightarrow{\rho }$\ the solution of the homogeneous BFKL
equation has the form (see \cite{conf})

\[
E_{\nu ,n}(\overrightarrow{\rho _{10}},\overrightarrow{\rho _{20}})\equiv
<\,\phi (\overrightarrow{\rho _{1}})\,O_{m,\widetilde{m}}(\overrightarrow{\rho
_{0}})\,\phi (\overrightarrow{\rho _{2}})\,>\,~=~\left( \frac{\rho _{12}}{%
\rho _{10}\rho _{20}}\right) ^{m}\left( \frac{\rho _{12}^{\ast }}{\rho
_{10}^{\ast }\rho _{20}^{\ast }}\right) ^{\widetilde{m}},
\]
where
\[
m=1/2+i\nu +n/2,\,\,~~\widetilde{m}=1/2+i\nu -n/2
\]
are the conformal weights simply related to eigenvalues of the Casimir
operators of the M\"{o}bius group. We use in $E_{\nu ,n}(\overrightarrow{%
\rho _{10}},\overrightarrow{\rho _{20}})$ the complex variables $\rho
_{k}=x_{k}+iy_{k}$ in the transverse subspace and the notation $\rho
_{kl}=\rho _{k}-\rho _{l}$.

For a principal series of the unitary representations the quantities $\nu $
and $n$ are respectively real and integer numbers. The projection $n$ of the
conformal spin $|n|$ can be positive or negative, but the eigenvalue of the
BFKL equation in LLA \cite{BL}
\bea
\omega =\omega ^{0}(n,\nu )~~=~~\frac{g^{2}N_{c}}{2\pi ^{2}}\left( \Psi
(1)-Re\,\Psi \left( \frac{1}{2}+i\nu +\frac{|n|}{2}\right) \right)
\label{2.2}
\eea
depends only on $|n|$. The M\"{o}bius invariance takes place also for the
Schr\"{o}dinger equation describing composite states of several reggeized
gluons \cite{BKP}.

It is important, that the above expression has the property of
the holomorphic separability \cite{separab}
\begin{equation}
\omega ^{0}(n,\nu )=\omega ^{0}(m)+\omega
^{0}(\widetilde{m})\,,\,\,\omega ^{0}(m)=\frac{g^{2}N_{c}}{8\pi
^{2}}\Bigl(2\Psi (1)-\Psi (m)-\Psi (1-m)\Bigr)\,  \label{separ}
\end{equation}
due to the relation
\[
\Psi (m)+\Psi (1-\widetilde{m})=\Psi (\widetilde{m})+\Psi (1-m)\,.
\]
Moreover, the more general property $H=h+h^{\ast }\,,\,\,[h,h^{\ast }]=0$ is
valid also for the Hamiltonian of an arbitrary number of reggeized gluons in
the multi-colour QCD $N_{c}\rightarrow \infty $ \cite{separab}. For the case
of the Odderon constructed from three gluons the holomorphic hamiltonian $h$
is integrable \cite{intodd}. Furthermore, in the multi-color QCD the BFKL
dynamics for an arbitrary number of reggeized gluons is also completely
integrable \cite{integr}. It is related to the fact, that in this case the
holomorphic Hamiltonian $h$ coincides with the local Hamiltonian for an
integrable Heisenberg spin model \cite{LFK}. The theory turns out to be
invariant under the duality transformation \cite{dual}. Presumably the
remarkable mathematical properties of the reggeon dynamics in LLA are
consequences of the extended $N=4$ supersymmetry \cite{N=4}, which is one of
the reasons for our investigation of the BFKL and DGLAP equations in this
model. It was argued in Ref. \cite{N=4} that generalized DGLAP equations for
matrix elements of quasi-partonic operators \cite{qp} for $N=4$ SUSY are
also integrable at large $N_{c}$. For the case of the Odderon a non-trivial
integral of motion discovered in Ref. \cite{intodd} was used to construct
the energy spectrum and wave functions of the three-gluon state \cite{WJ}. A
new Odderon solution with the intercept $j_{0}$ equal to 1 was found in
\cite
{bvl}. Recently the energy spectrum for reggeon composite states in the
multi-color QCD was obtained through the solution of the Baxter equation
(see refs. \cite{Vega} and \cite{Korch}). The effective action for reggeon
interactions is known \cite{Lipatov}. It gives a possibility to calculate
next-to-leading corrections for the gluon trajectory and reggeon couplings
\cite{FL}-\cite{KoLi}.

The solution of the inhomogeneous BFKL equation in the LLA approximation can
be written as the four-point Green function of a two-dimensional field
theory
\[
<\,\phi (\overrightarrow{\rho _{1}})\, \phi (\overrightarrow{\rho _{2}})\,
\phi (\overrightarrow{\rho _{1^{\prime }}})\, \phi (\overrightarrow{%
\rho_{2^{\prime }}})\,>\,=~ \sum_{n}\int_{-\infty }^{\infty }d\nu C(\nu
,|n|) \int d^{2}\rho _{0}\,\frac{E_{\nu ,|n|} (\overrightarrow{\rho_{10}},%
\overrightarrow{\rho _{20}}) E_{\nu ,|n|}^{\ast }(\overrightarrow{\rho
_{1^{\prime }0}}, \overrightarrow{\rho _{2^{\prime }0}})}{\omega -\omega
^{0}(|n|,\nu )},
\]
where $C(\nu ,|n|)$ is expressed through an inhomogeneous term of the
equation with the use of a completeness condition for $E_{\nu ,|n|}$ (see
\cite{conf}).

For $\overrightarrow{\rho _{1^{\prime }}}\rightarrow \overrightarrow{\rho
_{2^{\prime }}}$ we obtain $\overrightarrow{\rho _{01^{\prime }}}\sim
\overrightarrow{\rho _{1^{\prime }2^{\prime }}}$ and therefore
\begin{eqnarray}
<\,\phi (\overrightarrow{\rho _{1}})\,\phi (\overrightarrow{\rho _{2}})\,
\phi (\overrightarrow{\rho _{1^{\prime }}})\, \phi (\overrightarrow{%
\rho_{2^{\prime }}})\,>\, &\sim &\sum_{n}\int_{-\infty }^{\infty }d\nu \,
C(\nu ,|n|)\,\frac{E_{\nu ,|n|}(\overrightarrow{ \rho_{11^{\prime }}},%
\overrightarrow{\rho _{21^{\prime }}})} {\omega -\omega ^{0}(|n|,\nu )}\rho
_{1^{\prime }2^{\prime }}^{m}\rho _{1^{\prime }2^{\prime }}^{\ast 
\widetilde{m}}
\label{2.5} \\
&\sim &\sum_{n}C(\nu _{\omega },|n|)\,\frac{E_{\nu _{\omega },|n|} (%
\overrightarrow{\rho
_{11^{\prime }}},\overrightarrow{ \rho_{21^{\prime }}})%
}{\omega ^{\prime }(|n|,\nu _{\omega })} {|\rho _{1^{\prime }2^{\prime }}|}%
^{1+2i\nu _{\omega }} {\left( \frac{\rho _{1^{\prime }2^{\prime }}} {%
\rho_{1^{\prime }2^{\prime }}^{\ast }}\right) }^{|n|/2}\,,  \nonumber
\end{eqnarray}
where $\nu _{\omega }$ is a solution of the algebraic equation
\[
\omega =\omega ^{0}(|n|,\nu )
\]
with $Im \,\nu _{\omega }<0$.

The above asymptotics has a simple interpretation in terms of the Wilson
operator-product expansion
\[
\lim_{\rho _{1^{\prime }}\rightarrow \rho _{2^{\prime }}}\phi (%
\overrightarrow{\rho _{1^{\prime }}})\,\phi (\overrightarrow{\rho
_{2^{\prime }}})\,~=~\sum_{n}\,\frac{C(\nu _{\omega },|n|)}{\omega ^{\prime
}(|n|,\nu _{\omega })}{|\rho _{1^{\prime }2^{\prime }}|}^{2\Gamma
_{\omega }}%
{\left( \frac{\rho _{1^{\prime }2^{\prime }}}{\rho _{1^{\prime }2^{\prime
}}^{\ast }}\right) }^{|n|/2}O_{\nu _{\omega },|n|}(\overrightarrow{\rho
_{1^{\prime }}})\,,
\]
where
\[
\Gamma _{\omega }=\frac{1}{2}+i\nu _{\omega }
\]
is the transverse dimension of the operator $O_{\nu _{\omega },|n|}(%
\overrightarrow{\rho _{1^{\prime }}})$ calculated in units of the squared
mass. This operator is a mixed projection
\[
O_{\nu _{\omega },|n|}(\overrightarrow{\rho _{1^{\prime }}})~=~\widetilde{n}%
^{\mu _{1}}...\widetilde{n}^{\mu _{1+\omega }}\,l_{\bot }^{\sigma
_{1}}...l_{\bot }^{\sigma _{|n|}}\,O_{\mu _{1},...,\mu _{1+\omega },\sigma
_{1},...,\sigma _{|n|}}
\]
of a gauge-invariant tensor $O_{\mu _{1},...\mu _{J}}$ with $J=1+\omega
+|n|$
indices. Note, that because $\Gamma _{\omega }$ is real in the
deep-inelastic regime $\rho _{12}\rightarrow 0$ the operator $O_{\nu
_{\omega },|n|}(\overrightarrow{\rho _{1^{\prime }}})$ belongs to an
exceptional series of unitary representations of the M\"{o}bius group (see
\cite{vil}).

The anomalous dimension $\gamma (j)$ obtained from the BFKL equation in LLA
(see (\ref{2.2})) has the poles

\begin{equation}
\Gamma _{\omega }=1+\frac{|n|}{2}-\gamma (j),~~\gamma (j)|_{\omega
\rightarrow 0}=\frac{g^{2}N_{c}}{4\pi ^{2}\omega }\,.  \label{Gamma}
\end{equation}

Note, that there are also solutions of the BFKL equation corresponding to
operators with the shifted anomalous dimensions
\[
\Gamma _{\omega }(k)~=~\Gamma _{\omega }~+~k~~~(k=1,2,3,...).
\]
They contain the covariant laplacians $D_{\mu }^{2}$ and have higher twists.

The canonical contribution $1+|n|/2$ to the transverse dimension $\Gamma
_{\omega }$ corresponds to the local operator
\[
G_{\rho \mu _{1}}^{||}D_{\mu _{2}}^{||}...D_{\mu _{\omega }}^{||}D_{\sigma
_{1}}^{\bot }...D_{\sigma _{|n|}}^{\bot }G_{\rho \mu _{1+\omega }}^{||}
\]
in accordance with the fact, that in the light-cone gauge
$A_{\mu }\widetilde{n}%
_{\mu }=0$ the tensor

\[
G_{\rho \mu _{1}}G_{\rho \mu _{2}}\,\widetilde{n}^{\mu _{1}}\,\widetilde{n}^{\mu
_{2}}~\sim ~\partial _{\mu _{1}}A_{\rho }^{\bot }\partial _{\mu _{2}}A_{\rho
}^{\bot }\,\widetilde{n}^{\mu _{1}}\,\widetilde{n}^{\mu _{2}}
\]
has the transverse dimension equal to 1.

The local operator $O_{\nu _{\omega },|n|}$ for $|n|=1,\,2,...$ has the
twist higher than 2 because its anomalous dimension $\gamma $ is singular at
$\omega \rightarrow 0$. Indeed, for integer $|n|>0$ such singularity is
impossible for twist-2 operators, because in this case for $\omega
\rightarrow 0$ the total number of their indices $1+\omega +|n|$ would
coincide with the physical value of the Lorentz spin $j=1+|n|\geq 2$.
Moreover, the twist of $O_{\nu _{\omega },|n|}$ should grow at large $|n|$.
For the case $\omega >|n|>0$ this operator is a mixed projection of the
tensor
\[
G_{\rho \mu _{1}}G_{\mu _{2}\sigma _{1}}...G_{\mu _{|n|+1}\sigma
_{|n|}}D_{\mu _{|n|+2}}...D_{\mu _{\omega }}G_{\rho \mu _{1+\omega }}\,.
\]
In particular, for $n=0$ its twist equals 2. For $\omega <|n|$ the
corresponding operator is
\[
G_{\rho \mu _{1}}G_{\mu _{2}\sigma _{1}}...G_{\mu _{\omega }\sigma _{\omega
-1}}D_{\sigma _{\omega }}D_{\sigma _{\omega +1}}...D_{\sigma _{n}}G_{\rho
\mu _{1+\omega }}\,
\]
and its anomalous dimension has a singularity at $\omega \rightarrow 0$.

In the previous paper \cite{KoLi} we suggested to use an analytic
continuation of the BFKL anomalous dimension $\gamma (|n|,\omega )$ to the
points $|n|=-r-1$ ($r=0,1,2,...$) to calculate the singularities of the
anomalous dimension of the twist-2 operators at negative integer $j$
\begin{equation}
j=1+\omega +|n|\rightarrow -r\,.  \label{twist2}
\end{equation}
Because for fixed $|n|\neq -r-1$ and $\omega \rightarrow 0$ the quantity $%
\gamma (|n|,\omega )$ describes higher twist operators, to obtain the
anomalous dimension $\gamma $ for the twist-2 operators one should push
$|n|$
to $-r-1$ sufficiently rapidly at $\omega \rightarrow 0$

\begin{equation}
\Delta (|n|)\equiv |n|+r+1=\widetilde{C}_{1}(r)\,\omega ^{2}+\widetilde{C}%
_{2}(r)\,\omega ^{3}+...\,.  \label{Deltan}
\end{equation}
Note, that for $N=4$ SUSY there are several multiplicatively renormalizable
twist-2 operators (\ref{in4}) constructed from bosonic and fermionic fields.
Their anomalous dimensions can be obtained from the universal function $%
\gamma ^{uni}(j)$ (see Eq. (\ref{2})) by a shift of its argument $%
j\rightarrow j+k$ with $k=0,\,1,\,2,\,3,\,4$. It is related to the fact,
that these operators enter in the same supermultiplet. We assume in
accordance with Ref. \cite{KoLi}, that from the BFKL equation one is able to
calculate the singularities of $\gamma ^{uni}(j)$.

In LLA one can easily derive from the BFKL eigenvalue $\omega
^0(n,\nu)$ in the limit (\ref{twist2}) the following result for
$\gamma (j)$
\begin{equation} \gamma (j)|_{j\rightarrow
-r}~=~\frac{g^{2}N_{c}}{4\pi ^{2}}\frac{1}{j+r} \label{LLA}
\end{equation}
for all $r=-1,0,1,...\,$. Further, for the BFKL equation in LLA the fermions
are not important, but generally they give non-vanishing contributions to
the residues of the poles for $\gamma (j)$ in the DGLAP equation even in
LLA. Therefore the result (\ref{LLA}) for the anomalous dimension $\gamma
^{uni}(j)$ can be derived only for a certain generalization of QCD.

In our previous paper \cite{KoLi} it was shown, that only in an extended N=4
supersymmetric Yang-Mills theory the anomalous dimension, calculated in the
next-to-leading approximation from the BFKL equation, can be analytically
continued to negative $|n|$. Therefore the above result for $\gamma (j)$ in
LLA should be valid only for $N=4$ SUSY. Indeed, using the conservation of
the energy-momentum stress tensor $T_{\mu \nu }$ (corresponding to the
condition $\gamma (2)=0$) to fix a subtraction constant in the expansion of
$%
\gamma (j)$ over the poles at $j=-r$, we obtain :
\[
\gamma (j)~=~\frac{g^{2}N_{c}}{16\pi ^{2}}\gamma ^{LLA}(j),~~~\gamma
^{LLA}(j)~=~4\Bigl(\Psi (1)-\Psi (j-1)\Bigr)
\]
in an agreement with the direct calculation of $\gamma ^{uni}(j)$ in this
theory (see \cite{N=4,Dubna} and Section 5 below). It is important that this
anomalous dimension is the same for all twist-2 operators entering in the $%
N=4$ supermultiplet up to a shift of its argument by an integer number,
because this property leads to an integrability of the evolution equation
for matrix elements of quasi-partonic operators \cite{qp} in the
multi-colour limit $N_{c}\rightarrow \infty $ (see \cite{N=4}).\newline

In the NLO approximation there is a more complicated situation. Namely, the
anomalous dimension $\gamma ^{uni}(j)$ has the multiple poles $\Delta \gamma
\sim \alpha ^{2}(j+r)^{-3}$ at even $r$, which is related to an appearance
of the double-logarithmic corrections $\sim (\alpha \ln ^{2}s)^{n}$ in the
Regge limit $s\rightarrow \infty $ at upper orders $n$ of the perturbation
theory.

The origin of the double-logarithmic terms can be understood in a simple way
using as an example the process of the forward annihilation of the $%
e^{+}e^{-}$ pair in the $\mu ^{+}\mu ^{-}$ pair in QED \cite{FGGL}. Let us
write the scattering amplitude for this process in the following form

\[
f_{e^{+}e^{-}\rightarrow \mu ^{+}\mu ^{-}}~=4\pi \alpha _{em}\frac{\gamma
_{\sigma }^{\bot }\cdot \gamma _{\sigma }^{\bot }}{s}A(s,k_{\bot
}^{2})\,,~~~~\alpha _{em}=\frac{e_{em}^{2}}{4\pi }\,,\,\,s=-2p_{e}k\,,
\]
where $p_{e}$ and $-k$ are momenta of the incoming electron and positron
respectively and $k_{\bot }$ is the transverse part of $k$.
$A(s,k_{\perp })$
is the Bethe-Salpeter amplitude for the electron interaction with the
positron having the large virtuality $k_{\perp }^{2}\gg m^{2}$. The integral
equation for this amplitude in the double-logarithmic approximation ($\alpha
_{em}\ln ^{2}s\sim 1\,,\,\,\alpha _{em}\ll 1$) can be presented as follows
\cite{FGGL}
\bea
A(s,k_{\bot }^{2})~=~1~+~\frac{\alpha _{em}}{2\pi }\int_{m^{2}}^{s}
\frac{d{k^{\prime }}_{\bot }^{2}}{{k^{\prime }}_{\bot }^{2}}
\int_{{k^{\prime }}_{\bot }^{2}}^{\widetilde{s}}
\frac{ds^{\prime }}{s^{\prime }}A(s^{\prime },{k^{\prime }}_{\bot
}^{2})\,\,,
~~~~~~\left( \widetilde{s}=\min
\{s,s{k^{\prime }}%
_{\bot }^{2}/k_{\bot }^{2}\}\right) ,
\label{BSam}
\eea
where we took into account, that the double-logarithmic contribution appears
from the integration region

\[
s^{\prime }\ll s\,,\,\,\frac{{k^{\prime }}_{\bot }^{2}}{s^{\prime }}\gg
\frac{{k}_{\bot }^{2}}{s}\,,\,\,s^{\prime}\gg k_{\perp }^{\prime 2}\gg
m^{2}\,.
\]

One can search its solution in a form of the double Mellin transformation
\[
A(s,k_{\bot }^{2})~=~\int_{\sigma -i\infty }^{\sigma +i\infty }\frac{d\omega
}{2\pi i}{\left( \frac{s}{k_{\bot }^{2}}\right) }^{\omega }\int_{-i\infty
}^{i\infty }\frac{d\gamma }{2\pi i}{\left( \frac{k_{\bot }^{2}}{m^{2}}%
\right) }^{\gamma }f_{\omega }(\gamma )\,.
\]
Putting this anzatz in the equation we obtain a simple expression for 
$f_{\omega }(\gamma )$
\footnote{ The expression for $f_\omega (\gamma)$ in the published version
of this paper (see Nucl.Phys. B661 (2003) 19) was not  right because it did
not satisfy all boundary conditions contained in the integral equation (\ref{BSam}).
The correct expression (\ref{BSam1}) is derived in the Appendix D of
this hep-ph version and is sent by us as an Erratum to Nucl.Phys.B. 
We thank J. Bartels and M. Lyublinsky who drew our attention to this problem.}
\bea
f_{\omega }(\gamma )~=~\frac{(\omega -2\gamma )\,\gamma \,\omega \,
\lambda ^{-2}}{\omega -\frac{\alpha
_{em}}{2\pi }%
\,\left[ \frac{1}{\omega -\gamma }+\frac{1}{\gamma }\right] }\,.
\label{BSam1}
\eea
The pole of $f_{\omega }(\gamma )$ in $\gamma $ is situated at
\bea
\gamma =\frac{\omega }{2}\left( 1-\sqrt{1-\frac{2\alpha _{em}}{\pi \omega
^{2}}}\right) .
\label{BSam2}
\eea
Expanding this expression for the anomalous dimension $\gamma $ in the
series over $\alpha _{em}$ we shall generate the triple pole $\Delta \gamma
\sim \alpha _{em}^{2}/\omega ^{3}$.

It is important to note, that the pole contributions
\[
\frac{1}{\omega -\gamma }+\frac{1}{\gamma }
\]
in the denominator of $f_{\omega} (\gamma )$ are similar to singularities of
the expression
\begin{eqnarray}
\omega \sim 2\Psi (1)-\Psi (1+|n|+\omega -\gamma ) -\Psi (\gamma )
\label{db1}
\end{eqnarray}
(at $|n|=-1,-3,...$) appearing for $\omega \rightarrow 0$ in the eigenvalue
of the BFKL kernel of $N=4$ SUSY in a modified Born approximation (see Eqs.
(%
\ref{n7f}) and (\ref{n7g}) below). For odd negative $j$ the
double-logarithmic contributions are absent because the kernel of the
Bethe-Salpeter equation has an odd number of factors $k_{\bot }$ leading to
a cancellation of one logarithm after the integration of $\int d^{2}k_{\perp
}/k_{\perp }^{2}$ over the azimuthal angle $\varphi $.

It is obvious from the above QED example, that the Bethe-Salpeter equation
within the double-logarithmic accuracy can be derived both from the
BFKL-like and DGLAP-like equations by imposing more accurate constraints on
the region of integration. In the first case one should restrict $k_{\perp
}^{\prime 2}$ by $s^{\prime }$ from above, which results in the substitution
$-\gamma \rightarrow \omega -\gamma $ in the argument of the first $\Psi $%
-function (see the above expression for $\omega _{0}$). In the second case
one should restrict $s^{\prime }$ by $\widetilde{s}$ from above, which results
in the substitution $\gamma ^{-1}\rightarrow \gamma ^{-1}+(\omega -\gamma
)^{-1}$ and leads to a presence of the triple poles $\Delta \gamma \sim
\alpha ^{2}/\omega ^{3}$ in the two-loop anomalous dimension. This procedure
is similar to the rapidity veto approach invented by Bo Andersson and other
authors to explain the appearance of the large NLO corrections to the BFKL
kernel \cite{resum, Lund}.\newline


\section{NLO corrections to the BFKL kernel in $N=4$ SUSY}

\indent
To begin with, we review shortly the results of Ref. \cite{KoLi}, where NLO
corrections to the BFKL integral kernel at $t=0$ were calculated in the case
of QCD and supersymmetric gauge theories. Only $N=4$ SUSY is discussed in
details.\newline


\subsection{The set of eigenvalues}

The eigenvalues of the homogeneous BFKL equation for the $N=4$
supersymmetric gauge theory can be written as follows \cite{KoLi}
\begin{eqnarray}
\omega =4\,\overline{a}\,\biggl[\chi (n,\gamma )+\left( \frac{1}{3}\chi
(n,\gamma )+\delta (n,\gamma )\right) \,\overline{a}\,\biggr]\,,
\label{KL1}
\end{eqnarray}
where $\gamma =\frac{1}{2}-i\nu$ and
\begin{eqnarray}
\chi (n,\gamma ) &=&2\Psi (1)- \Psi (M)-\Psi (1-\widetilde{M}) \,,  \label{7} \\
&&  \nonumber \\
\delta (n,\gamma ) &=&\Psi ^{\prime \prime } (M) +\Psi ^{\prime \prime }
(1-%
\widetilde{M}) +6\zeta (3)-2\zeta (2)\chi (n,\gamma )  \nonumber \\
&-&2\Phi (|n|, \gamma )-2\Phi (|n|,1-\gamma )  \label{K3}
\end{eqnarray}
with
\[
M=\gamma +\frac{|n|}{2},~~~\widetilde M=\gamma -\frac{|n|}{2}
\]
Here 
$\Psi (z)$, $\Psi ^{\prime }(z)$ and $\Psi ^{\prime \prime }(z)$ are the
Euler $\Psi $ -function and its derivatives, respectively; $\overline{a}%
=g^{2}N_{c}/(16\pi ^{2})$ is expressed in terms of the coupling constant $g$
in the DREG scheme. The function $\Phi (|n|,\gamma )$ is given below
\footnote{%
Notice that the representation of $\Phi (|n|, \gamma )$ in terms of the sum
over $k$ in ref. \cite{KoLi} contained a misprint: the factors $(-1)^{k+1}$
in the last sum of (\ref{9}) were substituted by $(-1)^{k}$.}
\begin{eqnarray}
\Phi (|n|, \gamma ) &=&-\int_{0}^{1}dx~ \frac{x^{M -1}}{1+x}
\Biggr[\frac{1}{%
2}\biggl(\Psi ^{\prime }\Bigl(\frac{|n|+1}{2}\Bigr)- \zeta(2)\biggr)+ {\rm
Li%
}_{2}(-x)+{\rm Li}_{2}(x)  \nonumber \\
&+&\ln (x)\biggl(\Psi (|n|+1)-\Psi (1)+\ln (1+x)+
\sum_{k=1}^{\infty }\frac{%
(-x)^{k}}{k+|n|}\biggr)  \nonumber \\
&+&\sum_{k=1}^{\infty }\frac{x^{k}}{(k+|n|)^{2}}(1-(-1)^{k})\Biggr]
\nonumber \\
&&\hspace*{-1cm}\hspace{-1.3cm}=~\sum_{k=0}^{\infty }\frac{(-1)^{k+1}} {k+ M
}\Biggl[\Psi ^{\prime }(k+|n|+1)- \Psi ^{\prime }(k+1)+(-1)^{k+1}\Bigl(\beta
^{\prime }(k+|n|+1)+ \beta ^{\prime }(k+1)\Bigr) \biggr)  \nonumber \\
&-&\frac{1}{k+M }\biggl(\Psi (k+|n|+1)-\Psi (k+1)\biggr)\Biggr],  \label{9}
\end{eqnarray}
and
\begin{eqnarray}
\beta ^{\prime }(z)=\frac{1}{4}\Biggl[\Psi ^{\prime }\Bigl(\frac{z+1}{2}%
\Bigr)-\Psi ^{\prime }\Bigl(\frac{z}{2}\Bigr)\Biggr]= -\sum _{r=0}^{\infty}%
\frac{(-1)^r}{(z+r)^2}\,.  \label{9.0}
\end{eqnarray}

Note, that the term
\[
\frac{1}{3}\chi (n,\gamma )
\]
appears as a result of the use of the 
DREG scheme 
(see eq. (29) in \cite{KoLi}). It is well known, that DREG scheme violates
SUSY. The DREG procedure could be used provided that the numbers of boson
and fermion degrees of freedom would coincide (see \cite{CuCha}). But in its
standard version the number of degrees of freedom for vector fields is 
$4+2 \hat{\varepsilon}$ \footnote{%
Here $\hat{\varepsilon}=(D-4)/2$ and $D$ is the space-time dimension.} and
does not depend on $\hat{\varepsilon}$ for other particles. Usually (see
discussions in \cite{CaJoNi}-\cite{AvVla1}) it is possible to restore the
equality between boson and fermion degrees of freedom in SUSY by adding $2%
\hat{\varepsilon}$ additional scalar fields. 
Formally
it is equivalent to the dimensional reduction (for $N=4$ SUSY) from the
space-time dimension $10$ to the dimension $4+2\hat{\varepsilon}$ and we
call it below the DRED scheme. The change of the number of scalars from $6$
to $6-2\hat{\varepsilon}$ does not violate the cancellation of the singular
contribution $\sim \hat{\varepsilon}^{-1}$ to the coupling constant (the $%
\beta $-function is zero for $N=4$ SUSY in the DRED scheme \footnote{%
In the DRED scheme the $\beta $-function vanishing was demonstrated at first
three orders of perturbation theory (see \cite{AvVla,AvVla1} and references
therein).}) but leads to an additional contribution to $\omega $ of the
order of $\overline{a}^{2}$. Indeed, the correct result in the framework of
the DRED scheme can be obtained from eq.(\ref{KL1}) %
by a finite renormalization of the coupling constant
(see also G. Altarelli et al. and G.A. Shuler et al. in \cite{AvVla1})
\bea
{\overline{a}}\rightarrow \hat{a}={\overline{a}}+
\frac{1}{3}{\overline{a}}^{2},
\label{coco}
\eea
which eliminates the term $\sim \hat{a}^2 \chi $ in (\ref{KL1}).
For the new coupling constant $\hat{a}$ in the DRED scheme the
above expression for $\omega $ can be written in the following
simple form
\[ \omega =4\,\hat{a}\,\biggl[\chi (n,\gamma )+\delta
(n,\gamma )\,\hat{a}\, \biggr]\,.\,
\]



\subsection{Hermitian separability of the Bethe-Salpeter kernel}

Following the method of Ref. \cite{Dubna} and using the results of the
previous section we can write the above NLO correction to the BFKL equation
in the form having the property similar to the holomorphic separability (cf.
\cite{separab}). This property can be called the hermitian separability,
because it guarantees the reality of the eigenvalue $\omega $ for real $\nu
$%
.

Indeed, using the results of 
Appendix A we 
can obtain for the most complicated contribution to $\delta (n,\gamma )$
\begin{eqnarray}
&&\Phi (|n|,\gamma )+\Phi (|n|,1-\gamma )=\chi (n,\gamma )\,\left( \beta
^{\prime }(M)+\beta ^{\prime }(1-\widetilde{M})\right)   \nonumber \\
&&+\Phi _{2}(M)-\beta ^{\prime }(M)\left[ \Psi (1)-\Psi (M)\right] +\Phi
_{2}(1-\widetilde{M})-\beta ^{\prime }(1-\widetilde{M})\left[ \Psi (1)-\Psi (1-%
\widetilde{M})\right] ,  \nonumber
\end{eqnarray}
where $\chi (n,\gamma )$ is given by Eq.(\ref{7}).

Thus, we can rewrite the NLO corrections $\delta (n,\gamma )$ in a
(generalized) hermitian separable form (providing that $\omega _{0}$ is
substituted by $\omega $, which is valid with our accuracy):
\begin{eqnarray}
\delta (n,\gamma ) &=&\phi (M)+\phi (1-\widetilde{M})-\frac{\omega
_{0}}{2\hat{a}%
}\biggl(\rho (M)+\rho (1-\widetilde{M})\biggr),  \label{7.0} \\
\omega _{0} &=&4\hat{a}\biggl(2\Psi (1)-\Psi (M)-\Psi (1-\widetilde{M})\biggr)
\label{7.1}
\end{eqnarray}
and
\begin{eqnarray}
\rho (M) &=&\beta ^{\prime }(M)+\frac{1}{2}\zeta (2)\,,  \label{7.2} \\
\phi (M) &=&3\zeta (3)+\Psi ^{^{\prime \prime }}(M)-2\Phi _{2}(M)+2\beta
^{^{\prime }}(M)\Bigl(\Psi (1)-\Psi (M)\Bigr)\,.  \label{7.3}
\end{eqnarray}
In a simpler way the hermitian separability takes place in the eigenvalue
equation for the corresponding Bethe-Salpeter kernel
\begin{eqnarray}
1 &=&\frac{4\hat{a}}{\omega }\left( 2\Psi (1)-\Psi (M)-\Psi
(1-\widetilde{M}%
)+\hat{a}\left( \phi (M)+\phi (1-\widetilde{M})\right) \right)   \nonumber
\\
&&-2\hat{a}\left( \rho (M)+\rho (1-\widetilde{M})\right) .  \label{7.4}
\end{eqnarray}
In the right hand side the first contribution corresponds to its
singularity at $l=-1$ in the Gribov-Froissart representation generated by a
pole of the Legendre function $Q_{l}(z)$ at $\omega =1+l\rightarrow 0$ and
the last term appears from the regular part of the Born contribution. Note,
that $M$ and $1-\widetilde{M}$ coincide with the anomalous dimensions
appearing in the asymptotic expressions for the BFKL kernel in the limits
when the gluon virtualities are large: $q_{1}^{2}\rightarrow \infty $ and $%
q_{2}^{2}\rightarrow \infty $, respectively. Because $1-\widetilde{M}%
=M^{\ast }$, the hermitian separability guarantees the symmetry of $\omega $
for the principal series of the unitary representations of the
M\"{o}bius group (with $m=1/2+i\nu +n/2$) to the substitution 
$\nu \rightarrow -\nu $  and the
hermicity of the BFKL hamiltonian.


\subsection{The violation of a generalized holomorphical separability}

Eqs.(\ref{7.0})-(\ref{7.3}) show the hermitian separability with two
contributions depending on $M=\gamma +|n|/2$ and $M^{\ast }=1-\widetilde{M}%
=1-\gamma +|n|/2$. Although the above expressions are symmetric to $%
M\leftrightarrow M^{\ast }$, it is not clear, if we can symmetrize $\delta
(n,\gamma )$ to the substitution $|n|\leftrightarrow -|n|$ as it was done in
the Born approximation to obtain a true holomorphic separability (see Eq.(%
\ref{separ})). Using properties of polygamma-functions and Eqs. (\ref{6.91})
and (\ref{6.93}) one can attempt to present Eqs.(\ref{7.0})-(\ref{7.3})
in a separable form symmetric to the substitution $m\leftrightarrow
\widetilde{m}
$ (i.e., respectively, $M\leftrightarrow \widetilde{M}$ for both positive and
negative $n$). From the results of Appendix B the NLO correction $\delta
(n,\gamma )$ can be written in the form
\begin{eqnarray}
\delta (n,\gamma )
&=&\overline{\phi }(m)+\overline{\phi }(\widetilde{m})-\frac{%
\omega
_{0}}{2\hat{a}}\biggl(\overline{\rho }(m)+\overline{\rho }(\widetilde{m})%
\biggr)  \nonumber \\
&-&\pi ^{3}\left( \frac{\cos ^{2}(m\pi )}{\sin ^{3}(m\pi )}+\frac{\cos
^{2}(%
\widetilde{m}\pi )}{\sin ^{3}(\widetilde{m}\pi )}\right) +\delta
_{2}^{(2)}(n,\gamma ),  \label{holom1}
\end{eqnarray}
where
\begin{eqnarray}
\delta _{2}^{(2)}(n,\gamma ) &=&(1+(-1)^{n})\,\,\frac{\pi ^{2}\cos
(M\pi )}{%
\sin ^{2}(M\pi )}\Bigl[\Psi (M)-\Psi (\widetilde{M})\Bigr]\,  \nonumber \\
&=&(1+(-1)^{n})\,\varepsilon (n)\,\pi ^{2}\,\left( \frac{\Psi (m)\,\cos
(m\,\pi )}{\sin ^{2}(m\,\pi )}-\frac{\Psi (\widetilde{m})\,\cos
(\widetilde{m%
}\,\pi )}{\sin ^{2}(\widetilde{m}\,\pi )}\right) ~  \label{violat1}
\end{eqnarray}
with
\[
\varepsilon (n)=\left\{
\begin{array}{cc}
+1, & n\geq 0 \\
-1, & n<0
\end{array}
\right. \,.
\]

The term $\delta _{2}^{(2)}(n,\gamma )$ violates the holomorphic
separability. Because the contribution proportional to $\omega _{0}$ in
(\ref
{holom1}) depends not only on $M$ and $1-\widetilde{M}$ we have the
situation similar to the case of quantum anomalies. Namely, the property of
the hermitian separability of $\delta (n,\gamma )$ is responsible for the
violation of the holomorphic separability in $\delta _{2}^{(2)}(n,\gamma )$.
The anomalous term $\delta _{2}^{(2)}(n,\gamma )$ is zero for odd $n$, where
$\delta (n,\gamma )$ coincides with its analytic continuation to
corresponding negative $|n|$. Note, that for the Pomeron only even values of
$n$ are physical due to the Bose symmetry of its two-gluon wave function.
The colorless composite state of three reggeized gluons with the $f$%
-coupling in the color space and odd $n$ has an anti-symmetric wave function
and can give a large contribution to the small-$x$ behaviour of the
structure function $g_{2}(x)$ (see \cite{g2}).\newline



\subsection{Asymptotics of ``cross-sections'' at $s\to \infty$}

\indent
As an application of obtained results we consider the cross-section for the
inclusive production of two pairs of particles with their mass $\mu $ in the
polarized $\gamma \gamma $ collisions (see \cite{BL,KoLi}):
\[
\sigma (s)~=~\alpha _{em}^{2}\hat{a}^{2}
\frac{1}{{\mu }^{2}}\frac{32}{81}\biggl(\sigma _{0}(s)+\Bigl(\cos
^{2}\vartheta -\frac{1}{2}
\Bigr)\sigma _{2}(s)\biggr),
\]
where $\alpha _{em}$ is the electromagnetic fine structure constant, the
coefficient $\sigma _{0}(s)$ is proportional to the cross-section for the
interaction of unpolarized photons and $\sigma _{2}(s)$ describes the spin
correlation depending on the azimuthal angle $\vartheta $ between the
polarization vectors of colliding photons in their c.m system.

The asymptotic behavior of the cross-sections $\sigma _{k}(s)~~(k=0,2)$ at
$%
s\rightarrow \infty $ corresponds 
to an unmoving singularity of the $t$-channel partial wave $f_{\omega
}(t)\sim (\omega -\omega _{k})^{1/2}$ situated at
\begin{eqnarray}
\omega _{k} &=&4\hat{a}\Biggl[\chi \left( k,\frac{1}{2}\right) +\hat{a}%
\delta \left( k,\frac{1}{2}\right) \Biggr]\equiv 4\hat{a}\chi
\left( k,\frac{1}{2}\right) 
\Biggl[1-\hat{a}c\left( k,\frac{1}{2}\right) \Biggr]~~~\left(
c=-\frac{\delta }{\chi }\right) ,  \nonumber \\
&&  \nonumber \\
\sigma _{0}(s) &=&\frac{9\pi ^{5/2}}{32\sqrt{7\zeta (3)}}\frac{s^{\omega
_{0}}}{{\Bigl(\ln (s/s_{0})\Bigr)}^{1/2}}\cdot \biggl(1+O(\hat{a})\biggr),
\\
\sigma _{2}(s) &=&\frac{\pi ^{5/2}}{9\cdot 32\sqrt{7\zeta (3)-8}}\frac{%
s^{\omega _{2}}}{{\Bigl(\ln (s/s_{0})\Bigr)}^{1/2}}\cdot
\biggl(1+O(\hat{a})%
\biggr).
\end{eqnarray}
Here the symbol $O(\hat{a})$ denotes unknown next-to-leading corrections to
the impact factors. Note, that strictly speaking for the $N=4$
supersymmetric gauge theory we can not consider 
electro-magnetic interactions without the SUSY violation and it could be
more natural to investigate the interaction of gravitons or their
superpartners. But here we want only to illustrate the relative magnitudes
of the radiative corrections to the BFKL equation in QCD and in N=4 SUSY.

Using our results (\ref{7.0}) and (\ref{7.1}), we obtain in the $N=4$ case
the following numerical values for $\chi (k,1/2)$ and $c(k,1/2)$ $(k=0,2)$:
\begin{eqnarray}
\chi \left( 0,\frac{1}{2}\right)  &=&4\ln 2,~~~~~~~~\chi
\left( 2,\frac{1}{2}%
\right) ~=~4(\ln 2-1)\,,  \label{10} \\
&&  \nonumber \\
c\left( 0,\frac{1}{2}\right)  &=&2\zeta (2)+\frac{1}{2\ln 2}\Biggl[11\zeta
(3)-32{\rm Ls}_{3}\Big(\frac{\pi }{2}\Big)-14\pi \zeta
(2)\Biggr]~=~7.5812\,,
\label{11} \\
&&  \nonumber \\
c\left( 2,\frac{1}{2}\right)  &=&2\zeta (2)+\frac{1}{2(\ln 2-1)}\Biggl[%
11\zeta (3)+32{\rm Ls}_{3}\Big(\frac{\pi }{2}\Big)+14\pi \zeta (2)-32\ln 2%
\Biggr]  \nonumber \\
&=&6.0348\,,  \label{12}
\end{eqnarray}
where (see \cite{Lewin,Devoto}) 
\[
{\rm Ls}_{3}(x)=-\int_{0}^{x}\ln ^{2}\left| 2\sin \Bigl(\frac{y}{2}\Bigr%
)\right| dy\,.
\]
Note, that the function 
${\rm Ls}_{3}(x)$ appears also in contributions of some massive diagrams
(see, for example, the recent papers \cite{LS3} and references therein).

The LLA results (\ref{10}) coincide with ones obtained in ref. \cite{BL}. As
it was shown in \cite{FL,KoLi}, in the framework of QCD the NLO correction
$%
c^{QCD}(0,1/2)$ is large and leads to a strong reduction of the value of the
Pomeron intercept (see recent analyses \cite{bfklp,resum,L3,Lund} of various
resummations of the large NLO terms). Contrary to $c^{QCD}(0,1/2)$, the
correction $c(0,1/2)$ is not large ($c^{QCD}(0,1/2)/c(0,1/2) \approx 3.5$).
Because in N=4 SUSY the $\beta$-function is zero, it is natural to interpret
the large correction to the intercept of the BFKL Pomeron in QCD as an
effect related to the coupling constant renormalization. It seems to support
the results of refs. \cite{bfklp,DoShi} (see also the recent review \cite
{Lund} and references therein) concerning a large value for the argument of
the running QCD coupling constant at high energy processes.
The corrections $c^{QCD}(2,1/2)$ (see \cite{FL,KoLi}) and $c(2,1/2)$ for $%
N=4 $ are small and do not change significantly the LO value \cite{BL} of
the angle-dependent contribution.


\section{Non-symmetric choice of the energy normalization}

Analogously to 
refs. \cite{FL,KoLi} one can calculate eigenvalues of the BFKL kernel in the
case of a non-symmetric 
choice for the energy normalization $s_{0}$ in eq.(15) related to the
interpretation of the NLO corrections 
in the framework of the renormalization group (cf. \cite{FL,KoLi}). For the
scale $s_{0}=q^{2}$ natural for the deep-inelastic scattering we have
respectively

in DREG-scheme

\[
\omega =4\,\overline{a}\,\Biggl[\chi (n,\gamma -\frac{\omega }{2})+\biggl(%
\frac{1}{3}\chi (n,\gamma )+\delta (n,\gamma )\biggl)\,\overline{a}\,%
\Biggr]
\]
\[
=4\,\overline{a}\,\Biggl[\chi (n,\gamma )+\biggl(\frac{1}{3}\chi
(n,\gamma )+%
\widetilde{\delta}(n,\gamma )\biggl)\,\overline{a}\,\Biggr],
\]

in DRED-scheme
\[
\omega =4\,\hat{a}\,\Biggl[\chi (n,\gamma )+\widetilde{\delta}(n,\gamma )\,,%
\hat{a}\,\Biggr],
\]
where
\begin{eqnarray}
\widetilde{\delta}(n,\gamma )~=~\delta (n,\gamma )-2\chi (n,\gamma )\chi
^{\prime }(n,\gamma )
\label{tilddel}
\end{eqnarray}
and
\[
\chi ^{\prime }(n,\gamma )\equiv \frac{d}{d\gamma }\chi (n,\gamma )=-\Psi
^{\prime }\Bigl(M
\Bigr)+\Psi ^{\prime }\Bigl(1-\widetilde{M}
\Bigr)\,.
\]
Note, that here $\gamma $ does not coincide with the anomalous dimension of
the higher-twist operators with $|n|\geq 1$ (see below).


\subsection{Limit $\protect\gamma \to 0$ for $n=0$}

By considering the limit $\gamma \to 0$ of the BFKL eigenvalue
one can obtain for $n=0$ (see also the analysis in \cite{KoLi})
\begin{eqnarray}
\chi(0,\gamma ) &=& \frac{1}{\gamma} + O(\gamma^2), ~~  \nonumber \\
\frac{1}{3} \chi (n,\gamma ) &+& \widetilde \delta(0,\gamma ) ~=~
\frac{B^{DREG}%
}{\gamma} + C + O(\gamma^2),  \nonumber \\
\widetilde \delta(0,\gamma ) &=& \frac{B^{DRED}}{\gamma} + C + O(\gamma^2),
\label{n1}
\end{eqnarray}
where

\begin{eqnarray}
B^{DREG} ~=~ \frac{1}{3} ,~~~ B^{DRED} ~=~ 0, ~~~ C ~=~ 2\zeta(3).
\label{n7}
\end{eqnarray}

According to refs.\cite{FL,KoLi} with the use of eqs.(\ref{n1}) and
(\ref{n7}
) one can calculate the anomalous dimensions $\gamma$ of the twist-2
operators at $\omega \to 0$ (i.e. near $j=1$) respectively

in DREG-scheme
\begin{eqnarray}
\gamma = 4\, \overline{a}\, \Biggl[ \biggl( \frac{1}{\omega} + O(\omega)%
\biggr) + \overline{a}\, \biggl( \frac{ B^{DREG}}{\omega} + O(1) \biggr) +
\overline{a}^2 \, \biggl( \frac{C}{\omega^2} + O\left(\omega^{-1}\right) %
\biggr) \Biggl] ,  \label{n5}
\end{eqnarray}

in DRED-scheme
\begin{eqnarray}
\gamma = 4\, \hat a \, \Biggl[ \biggl( \frac{1}{\omega} + O(\omega)\biggr) +
\hat a \, \biggl( \frac{B^{DRED}}{\omega} + O(1) \biggr) + \hat a^2 \, %
\biggl( \frac{C}{\omega^2} + O\left(\omega^{-1}\right) \biggr) \Biggl].
\label{n6}
\end{eqnarray}

Thus, in the framework of the DRED scheme the singular contribution $\sim
\hat{a}^2 /\omega$ to the anomalous dimension is zero. These results will be
used below for the calculation of NLO corrections to $\gamma$ from the DGLAP
equation (see Eqs. (\ref{1a}) and (\ref{2})).\newline


\subsection{Hermitian separability for a non-symmetric normalization and the
symmetry $\protect\gamma \leftrightarrow J-\protect\gamma $}

For the scale $s_{0}=q^{2}$ the expression for $\omega $ as a function of
the anomalous dimension $\gamma =1/2+i\nu +|n|/2$ (correctly defined at
general $n$ (see (\ref{Gamma})) can be written in the following form
\[
\omega =4\hat{a}\biggl(\chi \left( n,\gamma -\frac{|n|}{2}\right) +\hat{a}%
\widetilde{\delta}\left( n,\gamma -\frac{|n|}{2}\right) \biggr),
\]
where $\chi (n,\gamma )$ and $\widetilde{\delta}(n,\gamma )$ (for the
non-symmetric choice of $s_{0}$) are given by Eqs. (\ref{7}), (\ref{K3}) and
(\ref{tilddel}), respectively.

By summing double-logarithmic terms in all orders of the perturbation theory
(see discussion in Section 2) we can write $\omega $ in the ''Lorentz
invariant'' form
\begin{eqnarray}
\omega =4\hat{a}\biggl(2\,\Psi (1)-\Psi (\gamma )-\Psi (1+|n|+\omega -\gamma
)+\widetilde{\Delta}(\,n,\,\gamma )\,\hat{a}\biggr)\,,
\label{n7d1}
\end{eqnarray}
where the Lorentz spin $j$ of the corresponding operators (for negative
$|n|$%
) is given by (\ref{twist2}). 
For $n=0$ the Lorentz spin of the twist-2 operator is $j=1+\omega $.

In the above expression
\begin{eqnarray}
\widetilde \Delta (\,n,\,\gamma )= \delta \left(n,\gamma-\frac{|n|}{2}\right)+
2\, \biggl[\Psi^{\prime }(\gamma )+\Psi^{\prime }(|n|+1-\gamma )\biggr] %
\chi\left(n,\gamma-\frac{|n|}{2}\right)\,.  \label{n7e}
\end{eqnarray}

Using the analysis of the subsection 3.2 one can present $\omega$
as follows
\begin{eqnarray}
\omega =4 \hat a\,\left( 2\,\Psi (1)-\Psi (\gamma )- \Psi (|n|+1+\omega
-\gamma)+\varepsilon \right) \,,\,\,  \label{n7f}
\end{eqnarray}
where $\varepsilon $ is written for $\omega \to 0$ in the form
\begin{eqnarray}
\varepsilon &=&\frac{\omega }2\Bigl( p (\gamma )+p (1+|n|-\gamma )\Bigr) +
\hat a\Bigl( \phi (\gamma )+\phi (1+|n|-\gamma )\Bigr) ,  \nonumber \\
\,\omega &=& 4 \hat a \Bigl( 2\,\Psi (1)-\Psi (\gamma )-\Psi (|n|+1-\gamma
)\Bigr) + O(\hat{a}^2)\,.  \label{n7g}
\end{eqnarray}

Here
\[
p (\gamma )~=~ \Psi ^{\prime }(\gamma )- \rho (\gamma ) ~=~ \Psi ^{\prime
}(\gamma ) -\beta ^{\prime }(\gamma ) -\frac{1}{2} \zeta(2)~=~2\sum_{k=0}^%
\infty \frac 1{(\gamma +2k)^2} -\frac{1}{2} \zeta(2)\,\,
\]
and $\phi (\gamma )$ is given by Eq.(\ref{7.3}).

In accordance with the hermitian separability of the BFKL kernel established
in the subsection 3.2 the eigenvalue equation for the corresponding
Bethe-Salpeter equation can be written as follows
\begin{eqnarray}
1 &=&\frac{4\hat{a}}{\omega }\left( 2\Psi (1)-\Psi (\gamma )-\Psi (J-\gamma
)+\hat{a}\left( \phi (\gamma )+\phi (J-\gamma )\right) \right)   \nonumber
\\
&&+2\hat{a}\left( p(\gamma )+p(J-\gamma )\right) \,,\,\,J=1+|n|+\omega \,,
\label{7.4}
\end{eqnarray}
where $J$ is the total number of tensor indices of the local operator which
does not coincide generally with its Lorentz spin $j$. Note that without the
shift $1+|n|-\gamma \rightarrow J-\gamma $ of the argument of $\Psi $%
-function at the LLA level, the symmetry of the eigenvalue to the
substitution $\gamma \rightarrow J-\gamma $ is violated by the term $\chi
^{\prime }(n,\gamma -|n|/2)$.\newline

Let us calculate $\widetilde \Delta (\,n,\,\gamma )$ near its singularities at
small $\gamma$. To begin with, we consider $\gamma \rightarrow 0$ for the
physical integer values $|n|\geq 0$:
\begin{equation}
\widetilde \Delta (\,n,\,\gamma ) \rightarrow \frac 4{\gamma ^2}\Bigl( \,\Psi
(1)- \Psi (|n|+1)\Bigr) \,+\frac 2\gamma \, \Bigl( c(n) +
2\Psi^{\prime}(|n|+1)\Bigr) \,,  \label{Delta}
\end{equation}
where

\begin{equation}
c(n)=\,\Psi ^{\prime }(|n|+1)-\Psi ^{\prime }(1)- \beta ^{\prime
}(|n|+1)+\beta^{\prime }(1)\,.  \label{c(n)}
\end{equation}

Therefore by solving the equation $\omega =\omega (n, \gamma)$ one can
obtain

\begin{equation}
\gamma =\frac{4\hat a}\omega \biggl( 1+\omega \left( \,\Psi (1)- \Psi
(|n|+1)\right) \biggr) +\frac{(4\hat a)^2}{\omega ^2} \left( \Psi (1)-\Psi
(|n|+1)+ \frac \omega 2c(n)\right) \,.  \label{gamma}
\end{equation}
At $n=0$ the correction $\sim \hat a$ to $\gamma $ is absent, but for other
$%
n$ we have the large correction

\[
\Delta \gamma =4\hat{a}\,\Bigl(\Psi (1)-\Psi (|n|+1)\Bigr)\,,
\]
having the poles at $1+\omega +|n|\rightarrow -r$, which leads to a
contribution changing even the singularities of the Born term. The
explanation of this effect is related to the presence of the
double-logarithmic terms $\Delta \gamma \sim \hat{a}^{2}/\omega ^{3}$ near
the points $j=0,2,...$ (see Sections 2 and 6). We remind, that for positive
integer $|n|$ we calculate the anomalous dimensions $\gamma $ of the higher
twist operators (with an anti-symmetrization between $n$ transversal and $%
1+\omega $ longitudinal indices). The singularities of the anomalous
dimensions of the twist-2 operators can be obtained only in the limit when
$%
\Delta (|n|)$ tends to zero more rapidly than $\omega $ (see Eq.
(\ref{Deltan}). \newline


\section{Anomalous dimension matrix in the $N=4$ SUSY}

The DGLAP evolution equation for the moments of parton distributions for $%
N=4 $ SUSY has the form
\begin{eqnarray}
\frac{d}{d\ln{Q^2}} f_a(j,Q^2) &=& \sum_{k} \gamma_{ab}(j) f_b(j,Q^2)
~~~~~~(a,b=q,g,\varphi),  \label{3.0} \\
\frac{d}{d\ln{Q^2}} \Delta f_a(j,Q^2) &=& \sum_{k} \widetilde \gamma_{ab}(j)
\Delta f_b(j,Q^2) ~~~~~~(a,b=q,g),  \label{3.01}
\end{eqnarray}
where the anomalous dimension matrices $\gamma_{ab}(j)$ and $\widetilde
\gamma_{ab}(j)$ can be obtained in the form of expansions over the coupling
constant $\hat a$ 
\begin{eqnarray}
\gamma_{ab}(j) &=& \hat a \cdot \gamma^{(0)}_{ab}(j) + \hat a^2 \cdot
\gamma^{(1)}_{ab}(j) + O(\hat a^3),~~~  \nonumber \\
\widetilde \gamma_{ab}(j) &=& \hat a \cdot \widetilde \gamma^{(0)}_{ab}(j) + \hat
a^2 \cdot \widetilde \gamma^{(1)}_{ab}(j) + O(\hat a^3).  \label{3.1}
\end{eqnarray}

In the following subsections we will present the results of exact
calculations for the leading order coefficients $\gamma _{ab}^{(0)}(j)$ and
$%
\widetilde{\gamma}_{ab}^{(0)}(j)$ and construct the anomalous dimensions of the
multiplicatively renormalizable twist-2 operators. In the NLO approximation
the corresponding coefficients $\gamma _{ab}^{(1)}(j)$ and $\widetilde{\gamma}%
_{ab}^{(1)}(j)$ were unknown (see, however, \cite{KoLiVe}). It is important,
that the form of the LLA anomalous dimension matrix of the multiplicatively
renormalizable operators in $N=4$ SUSY is rather simple because the result
is expressed in terms of one function. Taking into account this universality
related to the superconformal invariance, the existing information about the
anomalous dimensions of twist-2 operators in the QCD case, the known NLO
corrections to the BFKL kernel and an experience in integrating certain
types of the Feynman diagrams (see, for example, \cite{FleKoVe1,FleKoVe}),
we derive below the expressions for the NLO anomalous dimensions of the
multiplicatively renormalizable operators in $N=4$ SUSY. This result is
checked by direct calculations of the matrix elements $\gamma
_{ab}^{(1)}(j)$
and $\widetilde{\gamma}_{ab}^{(1)}(j)$ \cite{KoLiVe}.


\subsection{LLA results for the anomalous dimension matrix in $N=4$ SUSY}

The elements of the LLA anomalous dimension matrix in the $N=4$ SUSY have
the following form (see \cite{Dubna}):\newline

for tensor twist-2 operators
\begin{eqnarray}
\gamma^{(0)}_{gg}(j) &=& 4
\left( \Psi(1)-\Psi(j-1)-\frac{2}{j}+\frac{1}{j+1}
-\frac{1}{j+2} \right),  \nonumber \\
\gamma^{(0)}_{qg}(j) &=& 8 \left(\frac{1}{j}-\frac{2}{j+1}+\frac{2}{j+2}
\right),~~~~~~~~~~\, \gamma^{(0)}_{\varphi g}(j) ~=~ 12 \left( \frac{1}{j+1}
-\frac{1}{j+2} \right),  \nonumber \\
\gamma^{(0)}_{gq}(j) &=& 2 \left(\frac{2}{j-1}-\frac{2}{j}+\frac{1}{j+1}
\right),~~~~~~~~~~\, \gamma^{(0)}_{q\varphi}(j) ~=~ \frac{8}{j} \,,
\nonumber \\
\gamma^{(0)}_{qq}(j) &=& 4 \left( \Psi(1)-\Psi(j)+\frac{1}{j}-
\frac{2}{j+1}%
\right),~~ \gamma^{(0)}_{\varphi q}(j) ~=~ \frac{6}{j+1} \,,  \nonumber \\
\gamma^{(0)}_{\varphi \varphi}(j) &=& 4 \left( \Psi(1)-\Psi(j+1)\right),
~~~~~~~~~~~~~~~\, \gamma^{(0)}_{g\varphi}(j) ~=~ 4
\left(\frac{1}{j-1}-\frac{%
1}{j} \right),  \label{3.2}
\end{eqnarray}

for the pseudo-tensor operators:
\begin{eqnarray}
\widetilde \gamma^{(0)}_{gg}(j) &=& 4 \left( \Psi(1)-\Psi(j+1)-\frac{2}{j+1} +%
\frac{2}{j} \right),  \nonumber \\
\widetilde \gamma^{a,(0)}_{qg}(j) &=& 8 \left(-\frac{1}{j}+\frac{2}{j+1}\right),
~~~~~ \widetilde \gamma^{(0)}_{gq}(j) ~=~ 2\left( \frac{2}{j} -\frac{1}{j+1}%
\right),  \nonumber \\
\widetilde \gamma^{(0)}_{qq}(j) &=& 4 \left( \Psi(1)-\Psi(j+1)+\frac{1}{j+1}-
\frac{1}{j}\right).  \label{3.3}
\end{eqnarray}

Note, that in the $N=4$ SUSY multiplet there are also twist-2 operators with
fermion quantum numbers but their anomalous dimensions coincide up to an
integer shift of the argument with the above expressions for the bosonic
components (cf. ref. \cite{qp}).


\subsection{Anomalous dimensions and twist-2 operators with a multiplicative
renormalization}

It is possible to construct five independent twist-two operators with a
multiplicative renormalization. The corresponding parton distributions and
their LLA anomalous dimensions have the form (see \cite{Dubna}):

\begin{eqnarray}
f_I(j) &=& f_g(j) + f_{q}(j) + f_{\varphi}(j) \sim f^+_{q,g,\varphi}(j),
\nonumber \\
\gamma^{(0)}_{I}(j) &=& 4 \left( \Psi(1)-\Psi(j-1)\right) \equiv -4S_1(j-2)
\equiv \gamma^{(0)}_{+}(j),  \label{3.31} \\
& &  \nonumber \\
f_{II}(j) &=& -2(j-1)f_g(j) + f_{q}(j) + \frac{2}{3}(j+1) f_{\varphi}(j)
\sim f^0_{q,g,\varphi}(j),  \nonumber \\
\gamma^{(0)}_{II}(j) &=& 4 \left( \Psi(1)-\Psi(j+1)\right)\equiv -4S_1(j)
\equiv \gamma^{(0)}_{0}(j),  \label{3.32} \\
& &  \nonumber \\
f_{III}(j) &=& -\frac{j-1}{j+2}f_g(j) + f_{q}(j) - \frac{j+1}{j}
f_{\varphi}(j) \sim f^-_{q,g,\varphi}(j),  \nonumber \\
\gamma^{(0)}_{III}(j) &=& 4 \left( \Psi(1)-\Psi(j+3)\right)\equiv -4S_1(j+2)
\equiv \gamma^{(0)}_{-}(j),  \label{3.33} \\
& &  \nonumber \\
f_{IV}(j) &=& 2\Delta f_g(j) + \Delta f_{q}(j) \sim \Delta f^{+}_{q,g}(j),
\nonumber \\
\gamma^{(0)}_{IV}(j) &=& 4 \left( \Psi(1)-\Psi(j)\right)\equiv -4S_1(j-1)
\equiv \widetilde \gamma^{(0)}_{+}(j),  \label{3.34} \\
& &  \nonumber \\
f_{V}(j) &=& -(j-1)\Delta f_g(j) + \frac{j+2}{2} \Delta f_{q}(j) \sim \Delta
f^{-}_{q,g}(j),  \nonumber \\
\gamma^{(0)}_{V}(j) &=& 4 \left( \Psi(1)-\Psi(j+2)\right)\equiv -4S_1(j+1)
\equiv \widetilde \gamma^{(0)}_{-}(j),  \label{3.4}
\end{eqnarray}

Thus, we have one supermultiplet of operators with the same anomalous
dimension $\gamma ^{LLA}(j)$ proportional to $\Psi (1)-\Psi (j-1)$
\footnote{%
A similar result for another type of operators was obtained in ref. \cite
{DoOsb}.}. The momenta of the corresponding linear combinations of parton
distributions can be obtained from the above expressions $f_{k}(j)$ by an
appropriate shift of their argument $j$ 
in accordance with the corresponding shift of the argument of $\gamma
_{k}(j)
$. Moreover, the coefficients in these linear combinations for $N=4$ SUSY
can be found from the super-conformal invariance (cf. Ref \cite{qp}) and
should be the same for all orders of the perturbation theory in an
appropriate renormalization scheme.\newline

The momenta of three multiplicatively renormalizable twist-2 operators for
the unpolarized case are
\[
f_{N}(j)=a_{g}f_{g}(j)+a_{q}f_{q}(j)+a_{\varphi }f_{\varphi }(j)
\]
where the coefficients $a_{i}$ can be extracted from above expressions (\ref
{3.31})-(\ref{3.4}). If we insert this anzatz in the DGLAP equations (\ref
{3.0}) the following representations for the corresponding anomalous
dimensions
\begin{eqnarray}
\gamma _{N}^{(0)}(j) &=&\gamma _{gg}^{(0)}(j)+\frac{a_{q}}{a_{g}}\gamma
_{qg}^{(0)}(j)+\frac{a_{\varphi }}{a_{g}}\gamma _{\varphi g}^{(0)}(j)
\nonumber \\
&=&\gamma _{qq}^{(0)}(j)+\frac{a_{g}}{a_{q}}\gamma _{gq}^{(0)}(j)+\frac{%
a_{\varphi }}{a_{q}}\gamma _{\varphi q}^{(0)}(j)  \nonumber \\
&=&\gamma _{\varphi \varphi }^{(0)}(j)+\frac{a_{g}}{a_{\ps}}\gamma
_{g\varphi }^{(0)}(j)+\frac{a_{q}}{a_{\ps}}\gamma _{q\varphi }^{(0)}(j)
\label{fu.1}
\end{eqnarray}
can be obtained. Eqs.(\ref{fu.1}) lead to relations among the anomalous
dimension matrix $\gamma _{ab}^{(0)}(j)$ $(a,b=g,q,\varphi )$ which should
be valid also in the NLO approximation up to effects of breaking the
superconformal invariance (see \cite{KoLiVe} and references therein).

Analogously for the set of two multiplicatively renormalizable operators in
the polarized case
\begin{eqnarray}
\Delta f_N(j) &=& \widetilde a_g \Delta f_g(j) + \widetilde a_q \Delta f_{q}(j)
\nonumber
\end{eqnarray}
we can derive the following relations
\begin{eqnarray}
\widetilde \gamma^{(0)}_{N}(j) ~=~ \widetilde \gamma^{(0)}_{gg}(j) + \frac{\widetilde
a_q%
}{\widetilde a_g} \widetilde \gamma^{(0)}_{qg}(j) ~=~\widetilde \gamma^{(0)}_{qq}(j) +
\frac{\widetilde a_g}{\widetilde a_q} \widetilde \gamma^{(0)}_{gq}(j)\,.  \label{fu.2}
\end{eqnarray}

So, we have nine equations for the matrix elements in the case of the usual
partonic distributions and four equations for the polarized distributions,
which determines completely the anomalous dimension matrices $%
\gamma^{(0)}_{ab}(j)$ $(a,b=g,q,\varphi )$ and 
$\widetilde \gamma^{(0)}_{ab}(j)$
$(a,b=g,q )$ in terms of their eigenvalues in LLA
\[
\gamma^{(0)}_{\pm}(j) =-4S_1(j\mp2),~~~~ \gamma^{(0)}_{0}(j) =-4S_1(j) ,~~~~
\widetilde \gamma^{(0)}_{\pm}(j) =-4S_1(j\mp1)\,.
\]

This procedure is considered in details in Appendix C.



\subsection{NLO anomalous dimensions and twist-two operators with the
multiplicative renormalization}


We have the following initial information to predict the NLO anomalous
dimensions of twist-two operators with the multiplicative renormalization in
N=4 SUSY.\newline

{\bf 1.}~~ As it was shown in the previous subsections, the LLA anomalous
dimensions are meromorphic functions having the poles at
$j=-r,~r=-1,0,1,...$%
. Moreover, there is only one basic anomalous dimension $\gamma ^{LLA}(j)$
and all others can be obtained as $\gamma ^{LLA}(j+m)$, where $m$ is an
integer number. It is useful to choose (see Eq. (\ref{LLA})):

\[
\gamma ^{LLA}(j)=4\Bigl(\Psi (1)-\Psi (j-1)\Bigr)\equiv -4S_{1}(j-2).
\]
Then, $\gamma ^{LLA}(j)$ has a pole at $j\rightarrow 1$ and vanishes at
$j=2$%
. One should keep the above universality and linear relations among the
matrix elements also for the NLO anomalous dimensions $\gamma
_{ab}^{(1)}(j)$
and $\widetilde{\gamma}_{ab}^{(1)}(j)$ because it is a consequence of the
super-conformal invariance. It means, that we should construct only
the basic NLO anomalous dimension $\gamma ^{NLO}(j)$.\newline

{\bf 2.}~~ There are known results for the NLO corrections to the QCD
anomalous dimensions.\newline

{\bf 3.}~~ In the $\overline{MS}$-scheme with the coupling constant $%
\overline{a}$ and in the $\overline{MS}$-like scheme with the coupling
constant $\hat{a}$ (i.e. in the scheme based on DRED procedure), the terms
$%
\sim \zeta (2)$ should disappear in the final result 
for the forward Compton scattering (see \cite{CheKaTka}-\cite{Ko96}).
Therefore the terms $\sim \zeta (2)$ are cancelled at even $j$ in anomalous
dimensions for the structure functions $F_{2}$ and $F_{L}$ (related to the
unpolarized parton distributions) and at odd $j$ in anomalous dimensions for
structure functions $g_{1}$ and $F_{3}$ (related to the polarized parton
distributions).\newline

{\bf 4.}~~ From the BFKL equation in the framework of DRED scheme
(see  (\ref{7}), (\ref{K3}) and (\ref{9})) we know, that there is no
mixing among the functions of different transcendentality levels $i$
\footnote{%
Note that similar arguments were used also in \cite{FleKoVe} to obtain
analytic results for contributions of some complicated massive Feynman
diagrams. The method was based on a direct calculation of several terms in
the series over an inverse mass with taking into account its basic structure
found earlier in \cite{FleKoVe1,FleKoVe} by considering few special diagrams
with the use of the differential equation method \cite{DEM}.}, i.e. all
special functions at the NLO correction contain the sums of the terms $\sim
1/n^i ~(i=3)$. More precisely, if one will introduce the transcendentality
level of the functions in accordance with the complexity of the terms in the
corresponding sums
\[
\Psi \sim 1/n,~~~ \Psi^{\prime}\sim \beta^{\prime}\sim \zeta(2) \sim
1/n^2,~~~ \Psi^{\prime\prime}\sim \beta^{\prime\prime}\sim \zeta(3) \sim
1/n^3,
\]
then for the BFKL equation in LLA and in NLO the corresponding levels are $%
i=1$ and $i=3$, respectively.

Because in N=4 SUSY there is a relation between the BFKL and DGLAP
equations, the similar properties are assumed to be valid for anomalous
dimensions themselves, i.e. the basic functions $\gamma^{LLA}(j)$ and $%
\gamma^{NLO}(j)$ should be of the types $\sim 1/j^i$ with the levels $i=1$
and $i=3$, respectively. The only exception could be for the terms appearing
in the Born approximation, because such contribution can be removed by an
approximate finite renormalization of the coupling constant. The LLA basic
anomalous dimension is given above in terms of $S_1(j-2)$. Then,
the NLO basic anomalous dimension $\gamma^{NLO}(j)$ can be
expressed through the functions:
\[
S_{\pm i}(j-2),~ S_{\pm
k,l}(j-2), ~\zeta(k) S_{\pm l}(j-2),~ \zeta(i)
\]
(here $i=3$ and
$k+l=i$), where
\begin{eqnarray}
S_i(j)&=&\sum^{j}_{m=1}\frac{1}{m^i} \nonumber \\ S_{2,1}(j)&=&
\sum^{j}_{m=1}\frac{1}{m^2}\,S_1(m), \label{nl.0} \\
S_{-2}(j)&=&(-1)^j \sum^{j}_{m=1}\frac{(-1)^{m}}{m^2} -
\Bigl(1-(-1)^j \Bigr) \frac{1}{2}\,\zeta(2) ,  \nonumber \\
S_{-3}(j)&=&(-1)^j \sum^{j}_{m=1}\frac{(-1)^{m}}{m^3} - \Bigl(1-(-1)^j
\Bigr)%
\frac{3}{4}\,\zeta(3) ,  \nonumber \\
S_{-2,1}(j)&=&(-1)^j \sum^{j}_{m=1}\frac{(-1)^{m}}{m^2}\,S_1(m)
-\Bigl(1-(-1)^j \Bigr)\frac{5}{8}\,\zeta(3)  \label{nl.1}
\end{eqnarray}

The functions $S_{-2}(j)$, $S_{-3}(j)$ and $S_{-2,1}(j)$, which were
introduced recently in \cite{Verma}, 
coincide (up to an opposite sign) with the functions $K_{2}(j)$, $K_{3}(j)$
and $K_{2,1}(j)$ considered in \cite{KaKo,KaKo1} and used in our previous
paper \cite{KoLi01}, i.e. $S_{-i}(j)=-K_{i}(j)$, $S_{-2,1}(j)=-K_{2,1}(j)$.
Note, however, that our definition (\ref{nl.1}) of $S_{-i}(j)$ and $%
S_{-k,l}(j)$ coincides with the functions of \cite{Verma} only at even
values of $j$. Expressions 
(\ref{nl.1}) after the shift of the summation index $m\rightarrow m-j$ can
be analytically 
continuated 
from even to complex values of $j$, which reproduces $\beta ^{\prime }(j+1)$
(\ref{9.0}) 
and associated functions. 

Note that the terms $\sim \zeta (2)$ should be absent for even values of $j$
and for odd values of $j$ in the unpolarized and in polarized cases,
respectively, in accordance with item {\bf 3}.

Moreover, there is an important observation: the function
\[
S_{-1}(j) ~=~(-1)^j \sum^{j}_{m=1}\frac{(-1)^{m}}{m} - \Bigl(1-(-1)^j \Bigr)
\ln 2 
\]
does not contribute to the QCD anomalous dimensions and the Wilson
coefficient functions (see \cite{corAP} and \cite{KaKo,KaKo1},
respectively). Further, as it was shown in \cite{KaKo}, the terms $\sim
S_{-1}(j) $ cancel in the final results for the diagrams describing the
longitudinal Wilson coefficient function. \newline

{\bf 5.}~~ 
The terms
\begin{eqnarray}
S_{\pm l}(j-2)/(j \pm m)^k ~~~~(j+l=i)  \label{1.1a}
\end{eqnarray}
are absent in the anomalous dimension $\gamma^{NLO}(j)$. 
There are two reasons for this conclusion.

Firstly, these terms have additional poles in the points $j=\mp m$. But such
poles should cancel, if we start with the BFKL equation and obtain $\gamma
^{NLO}(j)$ by the analytic continuation to $|n|=-1-r$. Indeed, using such
procedure one can not obtain the doublets of poles.

The second reason comes from the consideration of the multiplicatively
renormalizable linear combinations (\ref{3.34}) and (\ref{3.4}). If, for
example, in the polarized case functions (\ref{1.1a}) contribute, then we
shall have the terms $\sim (j \pm m)^{-1}$ in one combination and the terms
$%
\sim (j \pm m \pm 2)^{-1}$ in another combination. However, from the direct
calculation of the NLO anomalous dimensions in the polarized case (see \cite
{MerNeer}) we know that only the terms $\sim j^{-1}$ and $\sim (j + 1)^{-1}$
are present in these combinations.

So, terms (\ref{1.1a}) should be absent in the universal NLO anomalous
dimension in the N=4 SUSY case.\newline

Note, that the sums $S_{-2}(j)$, $S_{-3}(j)$ and $S_{-2.1}(j)$
(see Eqs.(\ref{nl.1})) appear explicitly in calculations
only at even values of $j$ in the unpolarized case (and/or at odd values of
$%
j$, after the replacement $(-1)^{j}\rightarrow (-1)^{j+1}$, in the polarized
case). The analytic continuation to complex values of $j$ for the functions
$S_{\pm i}(j)$ ($i=1,2,3$) and $S_{-2.1}(j)$ in Eqs. (\ref{nl.0}) and (\ref
{nl.1}) can be done easily (see \cite{KaKo,AnalCont}) by a replacement of
the sums $\sum_{m=1}^{j}$ in the r.h.s with the difference $%
\sum_{m=1}^{\infty }-\sum_{m=1+j}^{\infty }$. They are expressed in terms of
the polygamma- and associated functions:
\begin{eqnarray}
S_{1}(j) &=&\Psi (j+1)-\Psi (1),  \nonumber \\
S_{i}(j) &=&\frac{(-1)^{i-1}}{(i-1)!}\biggl[\Psi ^{i-1}(j+1)-\Psi ^{i-1}(1)%
\biggr]=\frac{(-1)^{i-1}}{(i-1)!}\Psi ^{i-1}(j+1)+\zeta (i)~~~(i>1),
\nonumber \\
S_{2,1}(j) &=&2\zeta (3)-\sum_{m=1}^{\infty }\frac{S_{1}(m+j)}{(m+j)^{2}}%
=2\zeta (3)-\sum_{m=1}^{\infty }\frac{1}{(m+j)^{2}}\,\biggl[\Psi (j+1)-\Psi
(1)\biggr],  \label{nl.2} \\
S_{-1}(j) &=&\beta (j+1)-\beta (1)=\beta (j+1)+\ln 2\,,  \nonumber \\
S_{-i}(j) &=&\frac{(-1)^{i-1}}{(i-1)!}\biggl[\beta ^{i-1}(j+1)-\beta
^{i-1}(1)\biggr]=\frac{(-1)^{i-1}}{(i-1)!}\beta ^{i-1}(j+1)+\zeta
(-i)~~~(i>1),  \nonumber \\
S_{-2,1}(j)
&=&\sum_{m=0}^{\infty }\frac{(-1)^{m}S_{1}(m+j+1)}{(m+j+1)^{2}}-%
\frac{5}{8}\zeta (3)  \nonumber \\
&&\hspace{-1cm}=\sum_{m=0}^{\infty }\frac{(-1)^{m}}{(m+j+1)^{2}}\,\biggl[%
\Psi (m+j+2)-\Psi (1)\biggr]-\frac{5}{8}\zeta (3)~\equiv ~\widetilde{\beta}%
^{^{\prime \prime }}(j+1)-\frac{5}{8}\zeta (3),  \label{nl.3}
\end{eqnarray}
where $\zeta (-r)=\left( 2^{1-r}-1\right) \zeta (r)~~(r>1),~~~\zeta (-1)=\ln
2$.\newline


{\bf 6.}~~ Further, the NLO anomalous dimension $\gamma^{NLO}(j)$ is equal
to a combination of the most complicated contributions (i.e. the functions
with a maximal value of the transcendentality level $i=3$) for the QCD
anomalous dimensions (with the SUSY relation for the QCD color factors $%
C_F=C_A=N_c$). 

These most complicated contributions (with $i=3$) are the same for all QCD
anomalous dimensions (coinciding in N=4 SUSY) \cite{corAP,MerNeer} (only the
NLO scalar-scalar anomalous dimension is not known yet):
\begin{eqnarray}
\gamma^{(1)QCD}_{qq}(j)&\sim & \gamma^{(1)QCD}_{gg}(j) ~\sim ~ \widetilde
\gamma^{(1)QCD}_{qq}(j) ~\sim ~ \widetilde \gamma^{(1)QCD}_{gg}(j)  \nonumber \\
&=& 16 \,{\left(\frac{\alpha_s }{4\pi}\right)}^2 N_c^2\,\, Q(j-2) + ~...
~~~~~ (C_F=C_A=N_c)\,,  \label{1QCD}
\end{eqnarray}
where we omit less complicated contributions and
\begin{eqnarray}
Q(j) ~=~ \frac{1}{2} \biggl( S_3(j)+ S_{-3}(j)\biggr) +S_1(j) \biggl(%
S_2(j)+S_{-2}(j) \biggr) - S_{-2,1}(j)\,.  \label{1a}
\end{eqnarray}

Using Eqs. (\ref{nl.2}) and (\ref{nl.3}), the function $Q(j)$ can be
rewritten in terms of the polygamma- and associated functions:
\begin{eqnarray}
Q(j) &=&\frac{3}{4}\zeta (3)-\widetilde{\beta}^{^{\prime
\prime }}(j+1)+\frac{1}{%
4}\biggl(\Psi ^{^{\prime \prime }}(j+1)+\beta ^{^{\prime \prime }}(j+1)%
\biggr)  \nonumber \\
&+&\Bigl(\Psi (j+1)-\Psi (1)\Bigr)\biggl(\frac{1}{2}\zeta (2)-\Psi
^{^{\prime }}(j+1)-\beta ^{^{\prime }}(j+1)\biggr)  \nonumber \\
&=&\frac{3}{4}\zeta (3)-\widetilde{\beta}^{^{\prime \prime }}(j+1)+\frac{1}{16}%
\Psi ^{^{\prime \prime }}\left( \frac{j+1}{2}\right)   \nonumber \\
&+&\frac{1}{2}\Bigl(\Psi (j+1)-\Psi (1)\Bigr)\left( \zeta (2)-\Psi
^{^{\prime }}\left( \frac{j+1}{2}\right) \right) .  \label{1aa}
\end{eqnarray}

\vskip 0.5cm

Thus, for $N=4$ SUSY the NLO universal anomalous dimension $\gamma^{NLO}(j)$
has the form
\begin{eqnarray}
\gamma^{NLO}(j)~=~ 16 \, Q(j-2)  \label{1}
\end{eqnarray}

\vskip 0.5cm 

{\bf 7.}~~ We could add the term $\sim \zeta(3)$ to the r.h.s. of (\ref{1a}%
), but due to the condition\newline
$\gamma^{NLO}(j=2)=0$ it cancels. \newline

So, the universal anomalous dimension $\gamma(j)$ in two first orders of the
perturbation theory for N=4 SUSY is
\begin{eqnarray}
\gamma(j)~\equiv~ \gamma^{uni}(j)~=~ \hat a \gamma^{LLA}(j) + \hat a^2
\gamma^{NLO}(j), ~~~~~~~  \label{2}
\end{eqnarray}
where $\gamma^{LLA}(j)$ and $\gamma^{NLO}(j)$ are given by Eqs.
(\ref{LLA}) and (\ref{1}), respectively. All other anomalous
dimensions can be obtained as $\gamma^{uni}(j + m)$, where $m$ is
an integer number. \newline

Thus, the above arguments allow us to construct the NLO corrections to
anomalous dimensions in N=4 SUSY, which were unknown earlier. We check these
results by direct calculations \cite{KoLiVe} and reproduce the anomalous
dimension $\gamma ^{NLO}(j)$. Note, however, that in \cite{KoLiVe} 
the coupling constant $\overline{a}$ (see Eq.(\ref{coco}))
is used, which is responsable for an appearence
of the additional contribution $1/3\,\gamma ^{LO}(j)$ to the above obtained
NLO anomalous dimension (\ref{1}). This contribution can be absorbed in $%
\hat{a}$ by a finite renormalization of the coupling constant (see the
subsection 3.1).
Note also that in the dimensional reqularization scheme the eigenvalues
of this matrix are not expressed only in terms od the function
$ \gamma^{uni}(j)$ with the shift $j \to j+m$ of its argument.



\subsection{DGLAP evolution}

Using our knowledge of the anomalous dimensions we can construct the
solution of the DGLAP equation in the Mellin moment space in the framework
of N=4 SUSY.\newline

{\bf A.~~Polarized case}

The polarized parton distributions are splitted in two contributions:

\begin{eqnarray}
\Delta f_{q,g}(j,Q^2) ~=~ \Delta f^{+}_{q,g}(j,Q^2) + \Delta
f^{-}_{q,g}(j,Q^2),  \label{AP1}
\end{eqnarray}
where

at LO
\begin{eqnarray}
\Delta f^{\pm}_{q,g}(j,Q^2)~=~ \Delta f^{\pm,LO}_{q,g}(j,Q^2_0) {\Biggl(%
\frac{Q^2}{Q^2_0}\biggr)}^{\widetilde \gamma^{(0)}_{\pm}a} ~~~~~~~~~
\left(\widetilde \gamma^{(0)}_{\pm}=4S_1(j\mp 1) \right),  \label{AP2}
\end{eqnarray}

at NLO
\begin{eqnarray}
\Delta f^{\pm}_{q,g}(j,Q^2)~=~ \Delta f^{\pm,NLO}_{q,g}(j,Q^2_0) {\Biggl(%
\frac{Q^2}{Q^2_0}\biggr)} ^{\left(\widetilde \gamma^{(0)}_{\pm}a + \widetilde
\gamma^{(1)}_{\pm\pm}a^2\right)} ~~~~ \left(\widetilde \gamma^{(1)}_{\pm\pm}=16
Q(j\mp 1) \right),  \label{AP2}
\end{eqnarray}
where
\begin{eqnarray}
\Delta f^{\pm,NLO}_{q,g}(j,Q^2_0)&=& \Delta f^{\pm,NLO}_{q,g}(j,Q^2_0) %
\Biggl(1- \frac{\widetilde \gamma^{(1)}_{\pm\mp}\,\, \hat a} {\widetilde
\gamma^{(0)}_{\pm}- \widetilde\gamma^{(0)}_{\mp}} \Biggr)  \nonumber \\
&+& \frac{\widetilde \gamma^{(1)}_{\mp\pm}\,\, \hat a} {\widetilde
\gamma^{(0)}_{\mp}-\widetilde \gamma^{(0)}_{\pm}}\,\, \Delta
f^{\mp,NLO}_{q,g}(j,Q^2_0).  \label{AP3}
\end{eqnarray}
Here the anomalous dimensions $\widetilde \gamma^{(1)}_{\pm\mp}$ and $\widetilde
\gamma^{(1)}_{\mp\pm}$ are related to $\widetilde \gamma^{(1)}_{a,b}$
$(a,b=q,g)$
as follows:
\begin{eqnarray}
\left(
\begin{array}{cc}
\widetilde \gamma^{(1)}_{++}(j) & \widetilde \gamma^{(1)}_{-+}(j) \\
\widetilde \gamma^{(1)}_{+-}(j) & \widetilde \gamma^{(1)}_{--}(j)
\end{array}
\right) ~=~ {\hat{\widetilde V}}^{-1} \left(
\begin{array}{cc}
\widetilde \gamma^{(1)}_{gg}(j) & \widetilde \gamma^{(1)}_{qg}(j) \\
\widetilde \gamma^{(1)}_{gq}(j) & \widetilde \gamma^{(1)}_{qq}(j)
\end{array}
\right) \hat{\widetilde V}  \label{fu.5d}
\end{eqnarray}
with $\hat{\widetilde V}$ and ${\hat{\widetilde V}}^{-1}$ given in eq.(\ref{fu.6}).

Notice that only anomalous dimensions $\widetilde \gamma^{(1)}_{\pm\pm}$ are
important for $N=4$ SUSY at the order $O(\hat a^2)$ because they contribute
to the $Q^2$-evolution of parton distributions.

The anomalous dimensions $\widetilde{\gamma}_{\pm \mp }^{(1)}$, which were not
calculated in the previous section, can be found from eq.(\ref{fu.5d})
provided that the gluino-gluon polarized anomalous dimensions
$\widetilde{\gamma}%
_{ab}^{(1)}(j)$ $(a,b=g,q)$ are known. The anomalous dimensions $\widetilde{%
\gamma}_{\pm \mp }^{(1)}$ give contributions at the order $O(\hat{a}^{2})$
only to the normalization factors $\Delta f_{q,g}^{\pm ,NLO}(j,Q_{0}^{2})$.
They appear in the $Q^{2}$-dependent part of $\Delta
f_{q,g}^{\pm }(j,Q^{2})$
at the level $O(\hat{a}^{3})$ in the following form:
\[
\hat{a}^{3}\,\,\frac{\widetilde{\gamma}_{\pm \mp }^{(1)}\,\widetilde{\gamma}_{\mp
\pm }^{(1)}}{\widetilde{\gamma}_{\pm }^{(0)}-\widetilde{\gamma}_{\mp }^{(0)}}
\]

\vskip 0.5cm

{\bf B.~~Unpolarized case} 

The unpolarized parton distributions are splitted in the three parts:

\begin{eqnarray}
f_{q,g,\varphi}(j,Q^2) ~=~ \sum_{i=+,-,0} f^{i}_{q,g,\varphi}(j,Q^2),
\label{AP1}
\end{eqnarray}
where 

at LO
\begin{eqnarray}
f^{i}_{q,g,\varphi}(j,Q^2)~=~ f^{i,LO}_{q,g,\varphi}(j,Q^2_0)
{\Biggl(\frac{%
Q^2}{Q^2_0}\biggr)}^{\gamma^{(0)}_{i}a} ~~~~~~~~~
\left(\gamma^{(0)}_{\pm}=4S_1(j\mp2),~~\gamma^{(0)}_{0}=4S_1(j) \right),
\label{AP2}
\end{eqnarray}

at NLO
\begin{eqnarray}
f^{i}_{q,g,\varphi}(j,Q^2)~=~ f^{i,NLO}_{q,g,\varphi}(j,Q^2_0)
{\Biggl(\frac{%
Q^2}{Q^2_0}\biggr)} ^{\left(\gamma^{(0)}_{i}a + \gamma^{(1)}_{ii}a^2\right)}
\left(\gamma^{(1)}_{\pm\pm}=16 Q(j\mp2),\gamma^{(1)}_{00}=16 Q(j) \right)
\label{AP2}
\end{eqnarray}

As in the previous case {\bf A}, only the anomalous dimensions $\gamma
_{ii}^{(1)}$ are important at the level $O(\hat{a}^{2})$ in N=4 SUSY. The
anomalous dimensions $\gamma _{ik}^{(1)}$ $(i\neq k)$ contribute at $O(%
\hat{a}^{2})$ level only to the normalization factors $f_{q,g,\varphi
}^{i,NLO}(j,Q_{0}^{2})$. They are related to the qluino-gluon-scalar
anomalous dimensions $\gamma _{ab}^{(1)}(j)$ $(a,b=g,q,\varphi )$ as
follows:
\[
\left(
\begin{array}{ccc}
\gamma _{++}^{(1)}(j) & \gamma _{0+}^{(1)}(j) & \gamma _{-+}^{(1)}(j) \\
\gamma _{+0}^{(1)}(j) & \gamma _{00}^{(1)}(j) & \gamma _{-0}^{(1)}(j) \\
\gamma _{+-}^{(1)}(j) & \gamma _{0-}^{(1)}(j) & \gamma _{--}^{(1)}(j)
\end{array}
\right) ~=~{\hat{V}}^{-1}\left(
\begin{array}{ccc}
\gamma _{gg}^{(1)}(j) & \gamma _{qg}^{(1)}(j) & \gamma _{\varphi g}^{(1)}(j)
\\
\gamma _{gq}^{(1)}(j) & \gamma _{qq}^{(1)}(j) & \gamma _{\varphi q}^{(1)}(j)
\\
\gamma _{g\varphi }^{(1)}(j) & \gamma _{q\varphi }^{(1)}(j) & \gamma
_{\varphi \varphi }^{(1)}(j)
\end{array}
\right) \hat{V},~~
\]
where $\hat{V}$ and ${\hat{V}}^{-1}$ are given by eq.(\ref{fu.14}). The
anomalous dimensions $\gamma _{ik}^{(1)}$ $(i\neq k)$ can be obtained from
the above equation provided that the gluino-gluon-scalar
anomalous dimension matrix $\gamma _{ab}^{(1)}(j)$
$(a,b=g,q,\varphi )$ is known.


\section{Relation between the DGLAP and BFKL equations}

As we pointed out in subsection 3.6, in the case of $N=4$ SUSY the BFKL
results (\ref{7}) and (\ref{K3}) are analytic in $|n|$ and one can continue
the eigenvalues to the negative values of $|n|$. It gives a possibility to
find the singular contributions to anomalous dimensions of the twist-2
operators not only at $j=1$ but also at other integer non-physical points $%
j=0,\,-1,\,-2 ... $. In the Born approximation for the anomalous dimension
of the supermultiplet of the twist-2 operators we obtain $\gamma^{uni} = 4\,
\hat{a}\, 
(\Psi (1)-\Psi (j-1))$ which coincides with the result of the direct
calculations (see \cite{N=4,Dubna} and the discussions in subsection 4.2 ).
Thus, in the case of $N=4$ the BFKL equation presumably contains the
information sufficient for restoring the kernel of the DGLAP equation. Below
we investigate the relation between these equations in the NLO
approximation.


\subsection{DGLAP approach}

Let us start with an investigation of singularities of the anomalous
dimensions of twist-2 operators which were obtained in a direct way at the
previous Section.

By presenting the Lorentz spin $j$ as $\omega -r$, where $r=-1, 0,1,...$ and
pushing $\omega \rightarrow 0$ one can calculate the singular behavior of
the universal anomalous dimension $\gamma^{uni}(j) $ (\ref{2}). Note, that
in our discussion of the BFKL equation we used the more general definition
$%
j=1+\omega +|n|$ for the rank of the Lorentz tensor $O_{\mu _1...\mu _j}$
(see (\ref{twist2})), where $1+\omega$ is the number of its longitudinal
indices and $|n|$ is the conformal spin. Therefore strictly speaking in all
expressions discussed in this subsection $\omega$ should be substituted by
$%
j+r$.

For the special functions contributed to $\gamma ^{LLA}(j)$ and $\gamma
^{NLO}(j)$ at $j=\omega -r\rightarrow -r,r\geq 0$ we have the following
expansion in $\omega $:
\begin{eqnarray}
S_{1}(j-2) &=&-\frac{1}{\omega }+S_{1}(r+1)-\sum_{l=1}^{\infty }\omega ^{l}%
\biggl[(-1)^{l}\zeta (l+1)-S_{l+1}(r+1)\biggr]  \nonumber \\
S_{2}(j-2) &=&-\frac{1}{\omega ^{2}}-S_{2}(r+1)-\sum_{l=1}^{\infty
}(l+1)\omega ^{l}\biggl[(-1)^{l}\zeta (l+2)+S_{l+2}(r+1)\biggr]  \nonumber
\\
S_{3}(j-2) &=&-\frac{1}{\omega ^{3}}+S_{3}(r+1)  \nonumber \\
&-&\sum_{l=1}^{\infty }\frac{(l+1)(l+2)}{2}\omega ^{l}\biggl[(-1)^{l}\zeta
(l+3)-S_{l+3}(r+1)\biggr]  \label{dg2.1} \\
&&  \nonumber \\
S_{-2}(j-2) &=&(-1)^{r+1}\biggl\{\frac{1}{\omega ^{2}}+\zeta (-2)\left(
1-(-1)^{r}\right) +S_{-2}(r+1)  \nonumber \\
&+&\sum_{l=1}^{\infty }(l+1)\omega ^{l}\biggl[(-1)^{l}\zeta
(-(l+2))+S_{-(l+2)}(r+1)\biggr]\biggr\}  \nonumber \\
S_{-3}(j-2) &=&(-1)^{r+1}\biggl\{\frac{1}{\omega ^{3}}+\zeta (-3)\left(
1-(-1)^{r}\right) -S_{-3}(r+1)  \nonumber \\
&+&\sum_{l=1}^{\infty }\frac{(l+1)(l+2)}{2}\omega ^{l}\biggl[(-1)^{l}\zeta
(-(l+3))-S_{-(l+3)}(r+1)\biggr]\biggr\}  \nonumber \\
S_{-2,1}(j-2) &=&(-1)^{r+1}\biggl\{\frac{1}{\omega }\biggl[\zeta
(2)-S_{-2}(r+1)\biggr]  \nonumber \\
&&-\zeta (3)+\frac{5}{8}\zeta (3)\left( 1-(-1)^{r}\right)
-S_{-2,1}(r+1)-S_{-3}(r+1)\biggr\}~.  \label{dg2.3}
\end{eqnarray}

Presenting our results for $\gamma ^{LLA}(j)$ and $\gamma ^{NLO}(j)$ in
terms of these functions, we obtain
\begin{eqnarray}
\gamma ^{LLA}(j) &=& 4 \biggl[ \frac{1}{\omega} - S_1(r+1) - \hat S_{2}(r+1)
\omega + O(\omega^2) \biggr]  \label{dg2.4} \\
\gamma ^{NLO}(j) &=& 8 \biggl[ \frac{(1+(-1)^r)}{\omega^3} - 2S_1(r+1)
\frac{%
(1+(-1)^r)}{\omega^2}  \nonumber \\
&-& \Bigl((1+(-1)^r)\zeta(2) +2(-1)^r S_{2}(r+1) \Bigr) \frac{1}{\omega} +
O(\omega^0) \biggr]  \nonumber \\
& & \hspace{-2cm} = \Biggl\{
\begin{array}{cl}
16 \biggl[ \frac{1}{\omega^3} - 2S_1(r+1) \frac{1}{\omega^2} - \hat
S_{2}(r+1) \frac{1}{\omega} + O(\omega^0) \biggr] & \mbox{ if } r=2k \\
16 \biggl[ S_{2}(r+1) \frac{1}{\omega} + O(\omega^0) \biggr] & \mbox{ if }
r=2k+1\,,
\end{array}
\label{dg2.5}
\end{eqnarray}
where $\hat S_{2}(r) = \zeta(2) + S_{2}(r)$. So, the double-logarithmic
poles $\sim \omega ^{-3}$ appear in the case of even values of $r$ (see
discussions in Section 2).

Note that the functions $S_{-1}(r+1)$ and $S_{-2}(r+1)$ do not contribute to
$\gamma ^{LLA}(j)$ in the limit $j \to -r,~r \geq 0$. The absence of $%
S_{-1}(r+1)$ is explained by the fact, that the quantity $S_{-1}(j-2) $ does
not appear in the basic NLO anomalous dimension (see discussion in the
subsection 5.3, item 4).

On the other hand, the absence of the term $S_{-2}(r+1)$ is related to the
following important property of $\gamma$. As it follows from
eq.(\ref{dg2.3}%
), the functions $S_{-2}(j-2)$, $S_{-3}(j-2)$ and $S_{-2,1}(j-2)$ giving
contributions to $\gamma ^{NLO}(j)$ (see eqs. (\ref{1}) and (\ref{1a})) are
responsible for the different asymptotics of it at $j\rightarrow -r, r>0$
for even and odd $r$ (see the r.h.s. of eq.(\ref{dg2.5})). But the
combination of these functions $S_{-2}(j-2)$, $S_{-3}(j-2)$ and $%
S_{-2,1}(j-2)$ contributes to $\gamma ^{NLO}(j)$ in such way, that the
function $S_{-2}(r+1)$ is absent in the r.h.s. of eq.(\ref{dg2.5}).

Thus, for $\gamma ^{uni}(j)$ we obtain at $j=\omega -r\rightarrow -r$ ($%
r\geq 0$) \footnote{%
Because really the expansion parameter is $4\hat{a}$ (see, for example, Eqs.
(\ref{dg2.4}) and (\ref{dg2.5}), we shall consider below $G=4\hat{a}$ as a
coupling constant.}
\begin{eqnarray}
\gamma ^{uni}(j) &=&G\biggl[\frac{1}{\omega }-S_{1}(r+1)-\omega \,\hat{S}%
_{2}(r+1)+O(\omega ^{2})\biggr]  \nonumber \\
&+&G^{2}\Biggl\{
\begin{array}{cl}
\frac{1}{\omega ^{3}}-2S_{1}(r+1)\frac{1}{\omega
^{2}}-\hat{S}_{2}(r+1)\frac{%
1}{\omega }+O(\omega ^{0}) & \mbox{ if }r=2k \\
S_{2}(r+1)\frac{1}{\omega }+O(\omega ^{0}) & \mbox{ if }r=2k+1\,.
\end{array}
\label{dg2.59}
\end{eqnarray}

Note that $\gamma^{uni}(j)$ at $j =\omega -r \rightarrow -r$ can be
presented as a solution of the following Bethe-Salpeter equations
\begin{eqnarray}
1 &=& G \biggl[ \frac{1-\omega \, S_1(r+1)- \omega ^2\,\hat S_2 (r+1)}{%
\gamma (\omega-\gamma )} - S_1^2(r+1) \biggr]+\,...  \label{dg2.6}
\end{eqnarray}
for $r=2k \geq 0$ and
\begin{eqnarray}
1 &=& G \biggl[ \frac{1}{\gamma \omega}-\frac{S_1(r+1) +\omega \,\hat S_2
(r+1)}{\gamma} +S_2(r+1)\biggl]+\,...  \label{dg2.7}
\end{eqnarray}
for $r=2k+1>0$.

As it was shown in the subsection 5.4 the anomalous dimensions, responsible
for the $Q^2$ evolution of parton distributions in the framework of $N=4$
SUSY model, have the following form

in unpolarized case
\begin{eqnarray}
\gamma _{\pm}(j) &=& G \gamma^{(0)}_{\pm}(j) + G^2 \gamma^{(1)}_{\pm\pm}(j)
+ ... ~=~ \gamma^{uni}(j+2\mp2)  \nonumber \\
\gamma _{0}(j) &=& G \gamma^{(0)}_{0}(j) + G^2 \gamma^{(1)}_{00}(j) + ...
~=~ \gamma^{uni}(j+2)  \label{6.1n1}
\end{eqnarray}

and in polarized case
\begin{eqnarray}
\widetilde \gamma _{\pm}(j) &=& G \widetilde \gamma^{(0)}_{\pm}(j) + G^2 
\widetilde
\gamma^{(1)}_{\pm\pm}(j) + ... ~=~ \gamma^{uni}(j+1\mp2) \,.  \label{6.1n2}
\end{eqnarray}

Thus, for unpolarized and polarized cases, the corresponding anomalous
dimensions have the double-logarithmic poles $\omega ^{-3}$ at even and odd
$%
r$, respectively. The arguments of regular terms and of the functions in
front of poles $\omega ^{-i}$ are shifted by an integer number. By chosing
certain values of this number we obtain various $\gamma _s(j)$.


\subsection{BFKL approach}


Let us investigate a possibility to obtain the residues of the anomalous
dimension $\gamma ^{uni}(j)$ in the poles at $j=-r$ ($r=0,1,2,...$ ) from
the BFKL equation. Its eigenvalue $\omega $ can be analytically continued to
the negative points $|n|=-r-1$. One can present $\omega $ as a solution of
the equation (see (\ref{n7f}))
\[
\omega =G\left( 2\,\Psi (1)-\Psi (\gamma )-\Psi (|n|+1+\omega -\gamma
)+\varepsilon \right) \,,
\]
having the property of the hermitian separability corresponding to the
symmetry between $\gamma $ and $|n|+1+\omega -\gamma $ (see (\ref{7.4})).
The expression for $\varepsilon $ is given in Eq. (\ref{n7g}).

To begin with, we note that providing that $j=|n|+1+\omega$ (see Eq. (\ref
{twist2})) the r.h.s. of the BFKL equation contains the $\Psi$-function with
its argument equal to $j$ at $\gamma \to 0$, whereas $\gamma ^{LLA}(j)$
obtained from the DGLAP equation contains $\Psi (j-1)$.
The functions contributing to the NLO correction have an analogous shift of
their argument in comparison with the functions appearing in the NLO
corrections to $\gamma ^{uni}(j)$ (\ref{dg2.59}).

In the Born approximation $\gamma ^{LLA}\sim G$ the reason for the
difference of arguments of $\Psi $-functions appearing in the DGLAP and BFKL
approaches can be easily understood. Starting from the BFKL equation one
reproduces the universal anomalous dimension $\gamma ^{uni}$ of twist-two
operators by an analytic continuation of its eigenvalue $\omega$ to the
points $j=|n|+1+\omega =\omega -r$ ($r=0,1,...$). The residue of the pole of
$\gamma ^{uni}$ at $j=1$ is calculated also in accordance with the DGLAP
equation but without any continuation (see (\ref{n5}), (\ref{n6})). In the
BFKL approach the term regular at $j\rightarrow -r$ does not contain the
contribution from the pole at $j=1$ and equals $-GS_{1}(r)$ instead of $%
-GS_{1}(r+1)$. It is important, that the anomalous dimension obtained from
the BFKL equation depends on two parameters $\omega $ and $\left| n\right| $
and therefore we can impose a certain relation (see (\ref{Deltan})) between
two asymptotics $\omega \rightarrow 0$ and $\left| n\right| +r+1\rightarrow
0 $. Another ambiguity is related to the existence of double-logarithmic
terms near $j=-r$ (see below). As a result, in particular in LLA we can
obtain an agreement between the regular contribution to $\gamma ^{uni}$ and
its pole singularities.

In the next-to-leading approximation the situation is similar but not so
transparent. As in the case of LLA to obtain the residues of the poles for
$%
\gamma ^{uni}(j)$ in the BFKL approach, we continue the eigenvalue $\omega $
to $j=-r$ ($r=0,1,...$). But this continuation does not contain the regular
contributions from the singularity at $j=1$. Another problem is that to
calculate the residues of the non-leading poles $G^2\omega ^{-2}$ and $%
G^2\omega ^{-1}$ at $\omega \simeq j+r\rightarrow 0$ for $\gamma ^{uni}$ one
should continue the BFKL equation from the region of its applicapability $%
\omega \ll \gamma $ to the double-logarithmic region $\gamma \sim \omega $.
The property of the hermitian separability (\ref{7.4}) of the BFKL kernel is
helpful for this purpose, because it relates the small parameters in one
combination $\omega +|n|-r-1-\gamma $. It turns out, that in the
intermediate region $\gamma \sim \omega $ we should rearrange the BFKL
equation in such way that the Born contribution will contain the
double-logarithmic terms. This resummation procedure leads to a
renormalization of the next-to-leading correction $\sim G^{2}$. Namely, to
avoid the double-counting we should subtruct from this correction the sum of
poles containing the square of the most singular Born contribution. It turns
out, that the subtracted terms are small in the BFKL region $\omega \ll
\gamma $, but they are essential for the non-leading poles $G^2\omega ^{-2}$
and $G^2\omega ^{-1}$ of the anomalous dimension. These subtructions can be
found by calculating the next-to-leading contribution to the BFKL kernel
with a greater accuracy in comparison with ref. \cite{FL}, but we leave this
problem for future publications. Below we discuss in more details the origin
of above ambiguities.

Let us write down the eigenvalue of the BFKL kernel near the singularity at
$%
\gamma =0$ for the physical conformal spins $|n|=0,2,...$ (see Eqs.
(\ref{n7d1}), (\ref{n7e}) and 
(\ref{Delta}))

\begin{eqnarray}
\omega =G\left( \frac{1}{\gamma }-\Psi (1+|n|)+\Psi (1)\right) +G^{2}\left(
\frac{\Psi (1)-\Psi (1+|n|)}{\gamma ^{2}}+\frac{c(n)}{2\gamma }\right) \,.
\label{intn}
\end{eqnarray}
>From this equation one can obtain easily the anomalous dimensions (\ref
{gamma}) at integer positive $\left| n\right| $. The analytic continuation
of $\omega $ to the negative integer points $|n|\rightarrow -r-1$ contains
divergencies due to the following formulae

\[
\Psi (1)-\Psi (|n|+1)\rightarrow
\frac{1}{|n|+r+1}-S_{1}(r)-(|n|+r+1)\,\hat{S%
}_{2}(r)\,,
\]
\[
c(n)\rightarrow \frac{1+(-1)^{r}}{\left( |n|+r+1\right) ^{2}}%
+S_{2}(r)+(-1)^r S_{-2}(r)-\frac{1}{2} \left( 1+(-1)^{r}\right) \,\zeta
(2)\,,
\]
where $\hat{S}_{2}(r)$ was introduced after Eq. (\ref{dg2.5}).

We interprete the divergencies of $\omega $ at $|n|\rightarrow -r-1$ for
even $r=0,2,...$ as a manifestation of the double-logarithmic contribution
$%
\Delta \gamma \simeq G^{2}/\omega ^{3}$. Indeed, one can present expression
(%
\ref{intn}) in the kinematical region $\omega \ll \gamma \ll 1$ with the
same accuracy as a solution of the equation

\[
\omega =G\left( \frac{1}{\gamma }-\Psi (j-\gamma )+\Psi (1)\right)
+G^{2}\left( \frac{\Psi (1)-\Psi (j-\gamma )}{\gamma ^{2}}+\frac{%
c(j-1-\gamma )}{2\gamma }\right) ,
\]
where $j=1+\omega +\left| n\right| $. It is obvious, that if one neglects
the term $\sim G^{2}$ the anomalous dimension $\gamma $ contains the
double-logarithmic contribution $\Delta \gamma \sim G^{2}/\omega ^{3}$.
Note, that the above expression corresponds to a small-$\gamma $ asymptotics
of the BFKL equation written in the separable form (cf. (\ref{7.4}))

\begin{eqnarray}
1&=&G\left( \frac{2\Psi (1)-\Psi (\gamma )-\Psi
(j-\gamma )}{\omega }+\frac{%
c(\gamma -1 )+c(j-1-\gamma )}{2}+\zeta (2)\right)  \nonumber \\
&+&\frac{G^{2}}{4}\left( \phi (\gamma )+\phi (j-\gamma )\right) \,,
\nonumber
\end{eqnarray}
where $\phi (\gamma )$ is presented in Eq. (\ref{7.3}). On the other hand in
the Born approximation at $j\rightarrow -r$ and $\gamma \rightarrow 0$,
neglecting the corrections to the relation $\omega =j+r$, one can
simplify the previous equation 
as follows

\[
1\simeq G\left( \frac{1}{\gamma (\omega -\gamma )}-\frac{S_{1}(r)+(\omega
-\gamma )\,\hat{S}_{2}(r)}{\omega }\right) .
\]

It will be shown below, that for even $r$ the double-logarithmic terms $%
\Delta \gamma \sim G^{2}/\omega ^{3}$ obtained from this equation are not
cancelled. It is important, that the next-to-leading correction $\sim G^{2}$
to $\omega $ was obtained in ref. \cite{FL} by subtracting the first
iteration of the Born kernel from the contribution of all one-loop diagrams
to avoid the double-counting. At $\left| n\right| \rightarrow -r-1$ we chose
another Born approximation which contains the double-logarithmic terms. It
means, that now one should subtract from the next-to-leading correction some
terms including the most singular pole which appears in the first iteration
of this Born contribution to avoid the double-counting. But the subtracted
terms should be small in the applicapability region $\omega \ll \gamma $ for
the BFKL equation. Thus, the coefficient in front of $G^{2}$ after this
subtraction can be written for even $r$ as follows
\begin{eqnarray}
&&\frac{(\omega -\gamma )^{-1}-S_{1}(r)-(\omega -\gamma )\,\hat{S}_{2}(r)}{%
\omega \,\gamma ^{2}}+\frac{2(\omega -\gamma
)^{-2}+S_{2}(r)+S_{-2}(r)-2\zeta (2)}{2\omega \gamma }  \nonumber \\
&&-\frac{1}{(\omega -\gamma )^{2}\gamma ^{2}}+\frac{a(r)}{(\omega -\gamma
)\,\gamma ^{2}}+\frac{b(r)}{(\omega -\gamma )^{2}\gamma }\,+\frac{d(r)}{%
(\omega -\gamma )\,\gamma }+\frac{e(r)}{\gamma ^{2}}\,,  \nonumber
\end{eqnarray}
where the residues $a(r),b(r),d(r)\,$ and $e(r)$ are some unknown functions.
The second order pole $(\omega -\gamma )^{-2}$ and less singular
contributions are not essential with our accuracy. Thus, we obtain the
following modified BFKL equation at even $r$ and $\gamma \ll \omega $
\begin{eqnarray}
1=G\left( \frac{1}{\gamma \omega }+\frac{1}{\omega ^{2}}-\frac{%
S_{1}(r)+\omega \,\hat{S}_{2}(r)}{\omega }\right)+  \nonumber \\
G^{2}\left( \frac{a(r)-S_{1}(r)}{\gamma ^{2}\omega }+\frac{a(r)+b(r)}{\gamma
\omega ^{2}}+ \frac{S_{-2}(r)+3S_{2}(r) +\zeta (2)+2d(r)}{2\omega \gamma }+%
\frac{e(r)-\hat{S}_{2}(r)}{\gamma ^{2}}\right) .  \nonumber
\end{eqnarray}
If one will take into account, that $\omega $ and $j+r$ are generally
different parameters (see (\ref{Deltan})), it will be needed to substitute
in the above equation $\omega $ by $j+r$ and to change the unity at its left
hand side by an analytic function of $j+r$

\[
1\rightarrow 1+c_{1}(r)(j+r)+c_{2}(r)(j+r)^{2}+...\,\,,
\]
where $c_{1}(r)=-\widetilde{C}_{1}(r)$,
$c_{2}(r)=\widetilde{C}_{1}^{2}(r)-\widetilde{C}%
_{2}(r)$ and the parameters $\widetilde{C}_{1}(r)$ and $\widetilde{C}_{2}(r)$ 
are defined in Eq. (\ref{Deltan}).

After this modification of the equation we calculate the behavior of the
anomalous dimension at $j\rightarrow -r$ for even $r$ in terms of the
parameters $a(r),b(r),d(r),e(r)$ and $c_{1,2}(r)$
\begin{eqnarray}
\gamma =G\left( \frac{1}{j+r}+K(r)+L(r)\,(j+r)\right) +
G^{2}\left( \frac{1}{%
(j+r)^{3}}+\frac{T(r)}{(j+r)^{2}}+\frac{R(r)}{j+r}\right) \,,
\label{gamevj}
\end{eqnarray}
where
\begin{eqnarray}
K(r) &=&a(r)-S_{1}(r)-c_{1}(r)\,,~~T(r)=b(r)+K(r)-c_{1}(r)\,,  \nonumber \\
L(r) &=&e(r)-\hat{S}_{2}(r)-\Bigl(a(r)-S_{1}(r)%
\Bigr)^{2}+c_{1}^{2}(r)-c_{2}(r)\,,  \nonumber \\
R(r) &=&d(r)+\frac{S_{-2}(r)+S_{2}(r)-\zeta (2)}{2}-\left(
a(r)-S_{1}(r)\right) \left( b(r)+S_{1}(r)\right)  \nonumber \\
&-&c_{1}(r)(a(r)+b(r)-2S_{1}(r))+3c_{1}^{2}(r)-2c_{2}(r)\,.  \nonumber
\end{eqnarray}
This expression for $\gamma $ should be compared with the result (\ref
{dg2.59}) obtained in the previous subsection from the DGLAP equation. The
leading poles $G/(j+r)$ and $G^{2}/(j+r)^{3}$ coincide in both cases. As for
the regular terms $\sim G$ and residues of the poles $G^{2}/(j+r)^{2}$ and
$%
G^{2}/(j+r)$, the agreement can be achieved with an appropriate choice of
the parameters $a(r),b(r),d(r),e(r)$ and $c_{1,2}(r)$. Thus, to verify the
possibility, that the anomalous dimension $\gamma ^{uni}(j)$ can be derived
completely from the BFKL equation one should calculate these parameters from
the Feynman diagrams. As it is seen from the above expressions, the
parameters $c_{1,2}(r)$ enter together with $a(r),b(r),d(r),e(r)$ and could
be omitted. The vanishing of $c_{1,2}(r)$ corresponds to a natural
assumption, that to obtain the universal anomalous dimension of the twist-2
operators one should initially put $\left| n\right| =-r-1$ and only after
that $\omega $ can be pushed to zero. In this case we obtain $\gamma
^{uni}(j)$ (\ref{dg2.59}) if the parameters $a(r),b(r),d(r)$ are chosen as
follows
\begin{eqnarray}
a(r)
&=&-\frac{1}{r+1}\,,\,\,b(r)~=~-S_{1}(r+1)\,,\,\,e(r)~=~S_{1}^{2}(r+1)-%
\frac{1}{(r+1)^{2}}\,,  \nonumber \\
d(r) &=&\frac{S_{1}(r+1)}{r+1}+\frac{S_{-2}(r+1)-3S_{2}(r+1)}{2}\,.
\nonumber
\end{eqnarray}
As a result, the equation for the anomalous dimension can be written as
follows
\begin{eqnarray}
\hspace*{-1cm}1 &=&G\left( \frac{1}{\gamma (\omega -\gamma )}-\frac{%
S_{1}(r)+(\omega -\gamma )\,\hat{S}_{2}(r)}{\omega }\right) +  \nonumber \\
&&+G^{2}\Biggl(-\frac{S_{1}(r)}{\gamma ^{2}\omega }-\frac{(r+1)^{-1}}{\gamma
^{2}(\omega -\gamma )}-\frac{S_{1}(r+1)}{\gamma
(\omega -\gamma )^{2}}+\frac{%
S_{1}^{2}(r+1)-\hat{S}_{2}(r+1)}{\gamma ^{2}} \\
&&+\frac{S_{-2}(r)+3S_{2}(r)}{2\omega
\gamma }+\frac{\frac{2S_{1}(r+1)}{r+1}%
+S_{-2}(r+1)-3S_{2}(r+1)}{2\gamma (\omega -\gamma )}+ \frac{f(r)}{\gamma}%
\Biggr),  \nonumber
\end{eqnarray}
where the function $f(r)$ is fixed from the condition, that its solution
reproduces correctly the regular terms $\sim G^2$ in 
$\gamma ^{uni}(j)$ at $j \rightarrow -r$ (see (\ref{dg2.59})). 
Providing that we shall take into
account also non-singular terms at $\gamma \rightarrow 0$, the
BFKL  equation will contain an important information about
the singular part of the anomalous dimension in the order
$G^{3}$.

For odd $r$ the situation is simpler because here the double-logarithmic
terms at $j\rightarrow -r$ are absent. Indeed, in this case one can write
the BFKL equation in the form (after the subsequent substitution $\omega
\rightarrow j+r$)
\begin{eqnarray}
&&\hspace*{-1cm} 1+\widetilde{c}_{1}(r)\,\omega +\widetilde{c}%
_{2}(r)\,\,\omega ^{2}=G\left( \frac{1}{\gamma \omega }+\frac{1}{\omega
^{2}}%
-\frac{S_{1}(r)+\omega \,\hat{S}_{2}(r)}{\omega }\right) ~+~ G^{2} \Biggl(%
\frac{\widetilde{a}(r)-S_{1}(r)}{\gamma ^{2}\omega }  \nonumber \\
&&\hspace*{-1cm} -\frac{1}{\omega ^3 \gamma} +\frac{\widetilde{a}(r)+%
\widetilde{b}(r)}{\gamma \omega ^{2}}+\frac{-S_{-2}(r)+ 3S_{2}(r)+2\zeta
(2)+ 2\widetilde{d}(r)}{2\omega \gamma }\,+\, \frac{\widetilde{e}(r)\,-\,
\hat{S}_{2}(r)}{\gamma ^{2}} \Biggr) ,  \nonumber
\end{eqnarray}
where
$\widetilde{a}(r),\,\widetilde{b}(r),\,\widetilde{d}(r),\,\widetilde{e}%
(r)$\ and $\widetilde{c}_{1,2}(r)$ are the corresponding parameters for odd
$%
r$. Therefore the behavior of the anomalous dimension at $j\rightarrow -r$
for odd $r$ is
\begin{eqnarray}
\gamma &=& G\left( \frac{1}{j+r}+\widetilde{K}(r)+\widetilde{L}%
(r)\,(j+r)\right) +G^{2}\left( \frac{\widetilde{T}(r)}{(j+r)^{2}}+\frac{%
\widetilde{R}(r)}{j+r}\right) \,,  \label{gamodj} \\
\widetilde{K}(r)&=&\widetilde{a}(r)-S_{1}(r)-\widetilde{c}_{1}(r),~~
\widetilde{T}(r)~=~ \widetilde{b}(r)+2 \,\widetilde{K}(r) + \widetilde{c}%
_1(r),  \nonumber \\
\widetilde{L}(r)&=&\widetilde{e}(r)-\hat{S}_{2}(r)- (\widetilde{a}(r)
-S_{1}(r))^2 +\widetilde{c}_{1}^{2}(r)- \widetilde{c}_{2}(r) \,,  \nonumber
\\
\widetilde{R}(r)&=&\widetilde{d}(r)+ \widetilde{e}(r) - \frac{
S_{-2}(r)+S_{2}(r)+2\zeta(2)}{2}- \left(\widetilde{a}(r) -S_{1}(r)\right)
\left(3\widetilde{a} (r)+\widetilde{b}(r)-2S_1(r)\right)  \nonumber \\
&-&\widetilde{c}_{1}(r) \left(\widetilde{a}(r)+\widetilde{b}%
(r)-2S_1(r)\right) +2\widetilde{c}^2_1(r) -\widetilde{c}_{2}(r)\,.
\nonumber
\end{eqnarray}

Again we have a qualitative agreement with the result (\ref{dg2.59})
obtained from the DGLAP equation, but in the order $G^{2}$ the residues of
the poles $1/(j+r)^{2}$ and $1/(j+r)$ depends on parameters $\widetilde{a}%
(r),\,\widetilde{b}(r),\,\widetilde{d}(r),\widetilde{e}(r)\,$\ and $%
\widetilde{c}_{1,2}(r)$. If $\widetilde{c}_{1,2}(r)=0$, one can have a
complete agreement with Eq. (\ref{dg2.59}) provided that
\begin{eqnarray}
\widetilde{a}(r)
&=&-\frac{1}{r+1}\,,\,\,\widetilde{b}(r)=2S_{1}(r+1)\,,\,\,%
\widetilde{e}(r)=S_{1}^{2}(r+1)-\frac{1}{(r+1)^{2}}\,,  \nonumber \\
\widetilde{d}(r) &=&-S_{1}(r)S_{1}(r+1)+\frac{S_{-2}(r)-S_{2}(r)}{2}+\zeta
(2)+2S_{2}(r+1).  \nonumber
\end{eqnarray}
In this case the equation for the anomalous dimension can be written in the
form

\begin{eqnarray}
&&\hspace*{-1cm}1=G\left( \frac{1}{\gamma (\omega -\gamma )}-\frac{%
S_{1}(r)+(\omega -\gamma )\,\hat{S}_{2}(r)}{\omega }\right) ~+~  \nonumber
\\
&&G^{2}\Biggl( \frac{-1}{(\omega -\gamma )^{2}\omega \,\gamma }-\frac{%
S_{1}(r)}{\gamma ^{2}\omega }-\frac{(r+1)^{-1}}{\gamma
^{2}(\omega -\gamma )}%
+\frac{ 2S_{1}(r+1)}{\gamma \omega ^{2}}+\,\frac{S_{1}^{2}(r+1)-\,\hat{S}%
_{2}(r+1)}{ \gamma ^{2}} \\
&&\hspace*{-1cm}+\frac{3S_{2}(r)-S_{-2}(r)+2\zeta (2)}{2\omega \gamma }+%
\frac{\zeta (2)-S_{1}(r)S_{1}(r+1)+
\frac{S_{-2}(r)-S_{2}(r)}{2}+2S_{2}(r+1)%
}{\gamma (\omega -\gamma )}+\frac{\widetilde{f}(r)}{\gamma}\Biggr),
\nonumber
\end{eqnarray}
where $\widetilde{f}(r)$ is obtained from the condition that the regular
terms $\Delta \gamma \sim G^2$ at $j \rightarrow -r$ coincide with the
corresponding contributions to $\Delta \gamma ^{uni}$. Taking into account
also the non-singular contributions in the BFKL equation at
$\gamma \rightarrow 0$ one can obtain some information about
singularities of $\gamma $ in the order $G^{3}$.

Thus, the results for the universal anomalous dimension $\gamma ^{uni}(j)$
obtained from the BFKL equation with the use of our hypothesis, that $\gamma
^{uni}(j)$ coincides with $\gamma ^{BFKL}(j)$ after its analytic
continuation to the singular points $|n|=-r-1$, agree in main features with
the expressions derived directly from the DGLAP equation, but for a full
agreement one should verify by calculating the corresponding Feynman
diagrams, that the parameters $a(r),b(r),d(r),c_{1,2}(r)$ and
$\widetilde{a}%
(r),\,\widetilde{b}(r),\,\widetilde{d}(r),\widetilde{c}_{1,2}(r)$ reproduce
correctly the non-leading singularities $G^{2}/(j+r)^{2}$ and $G^{2}/(j+r)$.


\section{ Conclusion}

Above we reviewed the LLA results for the anomalous dimensions of twist-2
operators in the N=4 supersymmetric gauge theory and constructed the
operators with a multiplicative renormalization (cf. \cite{Dubna}). These
anomalous dimensions can be obtained from the universal anomalous dimension
$%
\gamma^{uni}(j)$ by a shift of its argument $j \to j+ k$ because they belong
to the same supermultiplet. We calculated the NLO correction to $\gamma
^{uni}(j)$ by using a number of plausible arguments. The results have a
compact form (see eqs. (\ref{1}) and (\ref{1a})) in terms of polylogarithms
and associated special functions. They are verified by direct calculations
\cite{KoLiVe}. Note that recently the LLA anomalous dimensions in this
theory were constructed in ref. \cite{GubKlePo} in the limit $j \to \infty$
from the superstring model with the use of the AdS/CFT correspondence \cite
{Malda,GubKlePo1,Witten}.

We investigated properties of the next-to-leading corrections to the kernel
of the BFKL equation in the N=4 supersymmetric theory. The absence of the
coupling constant renormalization in this model leads presumable to the
M\"{o}bius invariance of the BFKL equation in higher orders of the
perturbation theory. But the holomorphic separability of the BFKL kernel is
violated in the NLO approximation (see (\ref{holom})). Instead we have the
hermitial separability of the corresponding Bethe-Salpeter kernel (see (\ref
{7.4})). The cancellation of non-analytic contributions proportional to $%
\delta _{n}^{0}$ and $\delta _{n}^{2}$ in $N=4 $ SUSY is remarkable (such
terms contribute to $\omega$ in QCD and in N=1,2 supersymmetric models \cite
{KoLi,wu}). Moreover, the hermitian separability of the BFKL kernel is also
valid only in N=4 SUSY.

These properties could be a possible manifestation of the integrability of
the reggeon dynamics in the Maldacena model \cite{Malda} (see also \cite
{GubKlePo1,Witten}) which is reduced to the classical supergravity in the
limit $N_{c}\rightarrow \infty $. Indeed, in this N=4 supersymmetric model
the eigenvalues of the LLA pair kernels in the evolution equation for the
matrix elements of the quasi-partonic operators are proportional to $\Psi
(j-1)-\Psi (1)$ \cite{Dubna}, which means, that the corresponding
Hamiltonian at large $N_c$ coincides with the local Hamiltonian for an
integrable Heisenberg spin model \cite{N=4}. The residues of these
eigenvalues at the points $j=-r$ were obtained in Ref. \cite{KoLi} from the
BFKL equation in LLA by an analytic continuation of the anomalous dimensions
to negative integer values of the conformal spin $|n|$. It was shown above,
that the analogous correspondence between the BFKL and DGLAP equations takes
place qualitatively in the NLO approximation (cf. Eq. (\ref{dg2.59}) and
Eqs. (\ref{gamevj}, \ref{gamodj})). To verify quantitatively in this
approximation the hypothesis, that in N=4 SUSY the DGLAP equation can be
obtained from the BFKL equation, one should calculate the next-to-leading
corrections to the BFKL equation in the region, where $|n|+r+1 \rightarrow
0$
and $\gamma \sim \omega$. We hope to return to this problem in our future
publications.

\vspace{1cm} \hspace{1cm} {\Large {} {\bf Acknowledgments} 
}

The authors are supported in part by the INTAS 00-366 grant. A. Kotikov
thanks the Alexander von Humboldt Foundation and Landau-Heisenberg program
for a possibility to work on this problem in Germany. He was supported also
by the RFBR 02-02-17513 grant. L. Lipatov is thankful to the Hamburg,
Montpellier-2 and Paris-6 Universities for their hospitality during the
period of time when this work was in progress. He was supported also by the
RFBR and NATO grants.

We are indebted to V. Fadin, R. Kirschner, E. Kuraev, R. Peschansky, V.
Velizhanin and participants of the PNPI Winter School for helpful
discussions. We are grateful also to A. Belitsky, A. Gorsky and
G. Korchemsky for stimulation of discussion about DRED scheme and to
J. Bartels and M. Lyublinsky for their remarks concerning the double-logarithmic
behavior of the Bete-Salpeter amplitude $A(s,k_{\bot }^{2})$ in QED.

\section{Appendix A}

\label{App:A}
\def\theequation{A\arabic{equation}}
\setcounter{equation}0

Here we consider the most complicated contribution $\Phi (|n|,\gamma ) +
\Phi (|n|,1-\gamma )$ to the NLO correction $\delta (n,\gamma )$ in the BFKL
equation (\ref{K3}) and derive the property similar to the
holomorphic separability (cf. \cite{separab}). To begin with, let
us split the function $ \Phi (n,\gamma )$ in two terms

\[
\Phi (|n|,\gamma )=\Phi _{1}(|n|,\gamma )+\Phi _{2}(|n|,\gamma )\,,
\]
where
\[
\Phi _{1}(|n|,\gamma )=~\sum_{k=0}^{\infty }\frac{\left( \beta ^{\prime
}(k+|n|+1)-(-1)^{k}\Psi ^{\prime }(k+|n|+1)\right) }{k+M}
\]
\[
+\sum_{k=0}^{\infty }\frac{(-1)^{k}\Bigl(\Psi (|n|+k+1)-\Psi (1)\Bigr)}{%
(k+M)^{2}}\,
\]
and

\[
\Phi _2(|n|,\gamma )=~\sum_{k=0}^\infty \frac{\left( \beta ^{\prime
}(k+1)+(-1)^k\Psi ^{\prime }(k+1)\right) }{k+M }
\]
\[
-\sum_{k=0}^\infty \frac{(-1)^k\Bigl( \Psi (k+1)-\Psi (1)\Bigr) } {(k+
M)^2 }
\equiv \Phi _2(M )\,\,.
\]

It is convenient also to write the functions $\Phi _{1}(|n|,\gamma )$ and $%
\Phi _{2}(|n|,\gamma )$ as

\begin{eqnarray}
\Phi_1 (|n|,\gamma )&=&\Phi ^{(1)}(|n|,M)-\Phi ^{(2)}(|n|,M)\,,  \nonumber
\\
\Phi_2 (|n|,\gamma )&=&\Phi ^{(1)}(0,M)+\Phi ^{(2)}(0,M)\,,  \label{h.1}
\end{eqnarray}
where
\begin{eqnarray}
\Phi ^{(1)}(|n|,s)&=&\sum_{k=0}^\infty \frac{ \beta
^{\prime }(k+|n|+1) }{k+s%
} \,,  \label{h.2} \\
\Phi ^{(2)}(|n|,s)&=& \sum_{k=0}^\infty \frac{ (-1)^k}{k+s} \left( \Psi
^{\prime }(k+|n|+1) ~-~ \frac{ \Psi (|n|+k+1)-\Psi (1) }{k+s}\right)\,.
\label{h.3}
\end{eqnarray}

If we expand $\beta ^{\prime }(k+|n|+1)$ in the series
\[
\sum_{l=0}^{\infty }(-1)^{l+1}/(l+k+|n|+1)^{2}
\]
and insert it in the sum over $k$ in the r.h.s of Eq.(\ref{h.2}), we obtain

\[
\Phi ^{(1)}(|n|,s)~=~\Phi ^{(2)}(|n|,|n|-s+1)~+~ \beta ^{\prime }(|n|-s+1)%
\Bigl[\Psi (1)-\Psi (s)\Bigr]\,.
\]
In particular,
\begin{eqnarray}
\Phi ^{(1)}(|n|,M) &=&\Phi ^{(2)}(|n|,1-\widetilde M)~+~ \beta ^{\prime
}(1-\widetilde M)\Bigl[\Psi (1)-\Psi (M)\Bigr],  \nonumber \\
\Phi ^{(1)}(0,M) &=&\Phi ^{(2)}(0,1-M)~+~\beta ^{\prime }(1-M) \Bigl[\Psi
(1)-\Psi (M)\Bigr]\,.  \nonumber
\end{eqnarray}

The function $\Phi _{2}(|n|,\gamma )$ depends only on $M$ and therefore the
corresponding contribution to $\omega $ has the property of the hermitian
separability. 
Further, for $\Phi _{1}(|n|,\gamma )$ and $\Phi _{2}(|n|,\gamma )$ we derive
the following relation
\begin{eqnarray}
\Phi _{1}(|n|,\gamma ) &+&\Phi _{1}(|n|,1-\gamma )  \nonumber \\
&=&\Phi ^{(1)}(|n|,M)-\Phi ^{(2)}(|n|,M)+\Phi ^{(1)}(|n|,1-\widetilde{M})-\Phi
^{(2)}(|n|,1-\widetilde{M})  \nonumber \\
&=&\beta ^{\prime }(M)\left[ \Psi (1)-\Psi (1-\widetilde{M})\right] +\beta
^{\prime }(1-\widetilde{M})\left[ \Psi (1)-\Psi (M)\right] \,,  \label{6.91} \\
&&  \nonumber \\
\Phi _{2}(M) &=&\Phi ^{(1)}(0,M)+\Phi ^{(1)}(0,1-M)-\beta
^{\prime }(M)\left[
\Psi (1)-\Psi (1-M)\right] \,  \nonumber \\
&=&\Phi ^{(2)}(0,M)+\Phi ^{(2)}(0,1-M)+\beta ^{\prime }(1-M)\left[ \Psi
(1)-\Psi (M)\right] \,.  \label{6.93}
\end{eqnarray}
Therefore one can obtain
\begin{eqnarray}
&&\Phi (|n|,\gamma )+\Phi (|n|,1-\gamma )=\chi (n,\gamma )\,\left( \beta
^{\prime }(M)+\beta ^{\prime }(1-\widetilde{M})\right)   \nonumber \\
&&+\Phi _{2}(M)-\beta ^{\prime }(M)\left[ \Psi (1)-\Psi (M)\right] +\Phi
_{2}(1-\widetilde{M})-
\beta ^{\prime }(1-\widetilde{M})\left[ \Psi (1)-\Psi (1-%
\widetilde{M})\right] ,  \nonumber
\end{eqnarray}
where $\chi (n,\gamma )$ is given by Eq.(\ref{7}).\newline

\section{Appendix B}

\label{App:B}
\def\theequation{B\arabic{equation}}
\setcounter{equation}0

In this Appendix we demonstrate a violation of the generalized holomorhical
separability in the NLO correction $\delta (n,\gamma )$, i.e. we show an
impossibility to present Eqs.(\ref{7.0})-(\ref{7.3}) in the separable form
symmetric to the substitution $m\leftrightarrow \widetilde{m} $ (i.e.,
respectively, $M\leftrightarrow \widetilde{M}$ for both positive and negative 
$|n|$).

The expressions appearing in $\delta (n,\gamma )$ (\ref{K3}) can be written
as follows
\begin{eqnarray}
&&\hspace*{-1cm} \Psi ^{\prime \prime } (M) + \Psi ^{\prime \prime }
(1-\widetilde M) ~=~ \frac{1}{2} \biggl[ \Psi ^{\prime \prime } (M)+ 
\Psi^{\prime \prime } (1-M) + 
\Psi ^{\prime \prime } (\widetilde M) + \Psi ^{\prime \prime } 
(1-\widetilde M) \biggr],  \label{hs1} \\
&&\hspace*{-1cm} \Phi (|n|, \gamma )+ \Phi (|n|,1-\gamma ) ~=~ \Phi
^{(2)}(0,M)+\Phi ^{(2)}(0,1-M) + \Phi ^{(2)}(0,\widetilde M)+ \Phi
^{(2)}(0,1-\widetilde M)  \nonumber \\
&&\hspace*{-1cm} +\Bigl(\beta ^{\prime }(1-M) + \beta ^{\prime }(1-\widetilde
M)\Bigr) \left[ \Psi (1)-\Psi (M)\right] +\Bigl(\beta ^{\prime }(M) + \beta
^{\prime }(\widetilde M)\Bigr) \left[ \Psi (1)-\Psi (1-\widetilde M)\right].
\label{hs2}
\end{eqnarray}

Using the property
\[
\beta ^{\prime }(1-M) ~=~ \beta ^{\prime }(M) + \pi^2 \frac{\cos(M\pi)}{\sin
^2(M\pi)}\,,
\]
we can present the terms containing the functions $\beta ^{\prime }$
in the following form
\begin{eqnarray}
&&\beta ^{\prime }(1-M) + \beta ^{\prime }(1-\widetilde M) ~=~  \nonumber \\
&& \hspace*{-1cm} \frac{1}{2} \biggl[ \beta ^{\prime }(1-M) + \beta ^{\prime
}(1-\widetilde M) + 
\beta ^{\prime }(M) + \beta ^{\prime }(\widetilde M) + \pi^2
\frac{\cos(M\pi)}{\sin ^2(M\pi)} - \pi^2 \frac{\cos((1-\widetilde M)\pi)}{\sin
^2((1-\widetilde M)\pi)} \biggr] ,  \nonumber \\
&&\beta ^{\prime }(M) + \beta ^{\prime }(\widetilde M) ~=~  \nonumber \\
&& \hspace*{-1cm} \frac{1}{2} \biggl[ \beta ^{\prime }(M) + \beta ^{\prime
}(\widetilde M) + \beta ^{\prime }(1-M) + \beta ^{\prime }(1-\widetilde M) 
- \pi^2
\frac{\cos(M\pi)}{\sin ^2(M\pi)} + \pi^2 \frac{\cos((1-\widetilde M)\pi)}{\sin
^2((1-\widetilde M)\pi)} \biggr].  \nonumber
\end{eqnarray}

Then, the second line in Eq.(\ref{hs2}) can be replaced by
\begin{eqnarray}
&& \frac{1}{2} \biggl[ \beta ^{\prime }(M) + \beta ^{\prime }(\widetilde M) +
\beta ^{\prime }(1-M) + \beta ^{\prime }(1-\widetilde M)\biggr] ~\chi
(n,\gamma )
\nonumber \\
&&+ \frac{1}{2} \left( \frac{\pi^2\cos(M\pi)}{\sin^2(M\pi)} - \frac{%
\pi^2\cos((1-\widetilde{M})\pi)}{\sin^2((1-\widetilde{M})\pi)} \right) 
\Bigl[\Psi(1-%
\widetilde{M})-\Psi (M)\Bigr],  \label{hs3}
\end{eqnarray}
where $\chi (n,\gamma )$ from (\ref{KL1}) has the symmetric representation:
\begin{eqnarray}
\chi (n,\gamma ) &=&2\Psi (1)- \frac{1}{2} \biggl[ \Psi (M)+\Psi (1-M) +
\Psi (\widetilde M)+\Psi (1-\widetilde{M})\biggr] \,.  \label{hs4}
\end{eqnarray}

Thus, the NLO term $\delta (n,\gamma )$ is presented as follows
\begin{eqnarray}
\delta (n,\gamma )~=~ \delta_1 (n,\gamma ) + \delta_2 (n,\gamma ),
\nonumber
\end{eqnarray}
where
\begin{eqnarray}
\delta_1 (n,\gamma ) &=& \overline \phi(M) + \overline \phi(\widetilde{M}) -
\frac{\omega_0}{2\hat a} \biggl(\overline \rho(M) + \overline
\rho(\widetilde{M}%
) \biggr) ~~~~~~ \Bigl(\omega_0 =4\hat a \chi (n,\gamma )\Bigr),
\label{hs5}
\\
\delta_2 (n,\gamma ) &=& \left( \frac{\pi^2\cos(M\pi)}{\sin^2(M\pi)} -
\frac{%
\pi^2\cos((1-\widetilde{M})\pi)}{\sin^2((1-\widetilde{M})\pi)} \right) 
\Bigl[\Psi
(M)-\Psi (1-\widetilde{M})\Bigr]  \label{hs6}
\end{eqnarray}
and
\begin{eqnarray}
2\overline \rho (M) &=&\beta^{\prime}(M) + \beta^{\prime}(1-M)+ \zeta(2) \,,
\label{hs7} \\
2\overline \phi (M)&=& 6\zeta(3) + \Psi^{\prime\prime}(M) +
\Psi^{\prime\prime}(1-M) - 2\Phi^{(2)}(0,M) - 2\Phi^{(2)}(0,1-M)\,.
\label{hs8}
\end{eqnarray}

All above terms are symmetric to the substitution $M \leftrightarrow (1-%
\widetilde{M})$ in an agreement with previous results. Moreover, the term $%
\delta_1 (n,\gamma )$ is symmetric to $M \leftrightarrow \widetilde{M}$ (and,
thus, to $n \leftrightarrow -n$) due to the representations (\ref{hs7}) and
(%
\ref{hs8}). Therefore for this contribution the property of generalized
holomorphic separability is valid.

The term $\delta_2 (n,\gamma )$ can be presented as
\begin{eqnarray}
\left( \frac{\pi^2\cos(M\pi)}{\sin^2(M\pi)} +
\frac{\pi^2\cos(\widetilde{M}\pi)}{%
\sin^2(\widetilde{M}\pi)} \right) \Bigl[\Psi (M)-\Psi (\widetilde{M}) - 
\frac{%
\pi\cos(\widetilde{M}\pi)}{\sin(\widetilde{M}\pi)} \Bigr] \,,  \nonumber
\end{eqnarray}
where
\begin{eqnarray}
\frac{\pi\cos(\widetilde{M}\pi)}{\sin(\widetilde{M}\pi)} ~=~ 
\frac{\pi\cos(M\pi)}{%
\sin(M\pi)}\,.  \nonumber
\end{eqnarray}

It can be splitted in two parts: $\delta_2 (n,\gamma )=\delta_2^{(1)}
(n,\gamma )+\delta_2^{(2)} (n,\gamma )$ where the contribution
\begin{eqnarray}
\delta_2^{(1)} (n,\gamma ) &=& -\biggl( \frac{\pi^2\cos(M\pi)}{\sin^2(M\pi)}
+ \frac{\pi^2\cos(\widetilde{M}\pi)}{\sin^2(\widetilde{M}\pi)} \biggr)
\frac{\pi\cos(%
\widetilde{M}\pi)}{\sin(\widetilde{M}\pi)}  \nonumber \\
&=& -\frac{\pi^3\cos^2(M\pi)}{\sin^3(M\pi)} -
\frac{\pi^3\cos^2(\widetilde{M}\pi)%
}{\sin^3(\widetilde{M}\pi)}  \nonumber
\end{eqnarray}
is symmetric under the substitution $M \leftrightarrow \widetilde{M}$ (and,
thus, to $n \leftrightarrow -n$). So, the term $\delta_2^{(1)} (n,\gamma )$
has also the property of the generalized holomorphic separability.

Thus, the sum of the contributions $\delta_1 (n,\gamma )$ and $%
\delta_2^{(1)} (n,\gamma )$ has the generalized separable form symmetric to
the substitution $m \leftrightarrow \widetilde{m}$:
\begin{eqnarray}
\delta_1 (n,\gamma ) + \delta_2^{(1)} (n,\gamma ) &=& \overline \phi(m) +
\overline \phi(\widetilde m) - 
\frac{\omega_0}{2\hat a} \biggl(\overline \rho(m)
+ \overline \rho(\widetilde m) \biggr)  \nonumber \\
&-& \pi^3\left(\frac{\cos ^2(m\pi)}{\sin ^3(m\pi)} +\frac{\cos
^2(\widetilde{%
m}\pi)}{\sin ^3(\widetilde{m}\pi)}\right)  \label{h7.0}
\end{eqnarray}

The last term
\begin{eqnarray}
\delta_2^{(2)} (n,\gamma ) ~=~ \left( \frac{\pi^2\cos(M\pi)}{\sin^2(M\pi)} +
\frac{\pi^2\cos(\widetilde{M}\pi)}{\sin^2(\widetilde{M}\pi)} \right) 
\Bigl[\Psi
(M)-\Psi (\widetilde{M}) \Bigr],  \nonumber
\end{eqnarray}
is anti-symmetric under the transformation $M \leftrightarrow \widetilde{M}$,
i.e.

when $n>0$, and thus, $M=m,~~\widetilde{M}=\widetilde{m}$~ it equals to
\begin{eqnarray}
\left( \frac{\pi^2\cos(m\pi)}{\sin^2(m\pi)} +
\frac{\pi^2\cos(\widetilde{m}\pi)}{%
\sin^2(\widetilde{m}\pi)} \right) \Bigl[\Psi (m)-\Psi (\widetilde{m}) \Bigr]
\nonumber
\end{eqnarray}


and when $n<0$, and thus, $M=\widetilde{m},~~\widetilde{M}=m$~ it is
\begin{eqnarray}
\left( \frac{\pi^2\cos(m\pi)}{\sin^2(m\pi)} +
\frac{\pi^2\cos(\widetilde{m}\pi)}{%
\sin^2(\widetilde{m}\pi)} \right) \Bigl[\Psi (\widetilde{m})-\Psi (m) \Bigr].
\nonumber
\end{eqnarray}

Therefore the term $\delta _{2}^{(2)}(n,\gamma )$ in the next-to-leading
correction
\begin{eqnarray}
\delta (n,\gamma )
&=&\overline{\phi }(m)+\overline{\phi }(\widetilde{m})-\frac{%
\omega
_{0}}{2\hat{a}}\biggl(\overline{\rho }(m)+\overline{\rho }(\widetilde{m})%
\biggr)  \nonumber \\
&-&\pi ^{3}\left( \frac{\cos ^{2}(m\pi )}{\sin ^{3}(m\pi )}+\frac{\cos
^{2}(%
\widetilde{m}\pi )}{\sin ^{3}(\widetilde{m}\pi )}\right) +\delta
_{2}^{(2)}(n,\gamma )  \label{holom}
\end{eqnarray}
violates the holomorphic separability.
Note, that the anomalous term $\delta_{2}^{(2)}(n,\gamma )$ can be written
as Eq. (\ref{violat1}), 
i.e. it is zero for odd $n$, where $\delta (n,\gamma )$ coincides with its
analytic continuation to corresponding negative $|n|$.

\section{Appendix C}

\label{App:C}
\def\theequation{C\arabic{equation}}
\setcounter{equation}0

Here we construct the anomalous dimension matrices $\gamma _{ab}^{(0)}(j)$
$%
(a,b=g,q,\varphi )$ and $\widetilde{\gamma}_{ab}^{(0)}(j)$ $(a,b=g,q)$ in 
terms of their eigenvalues in LLA
\[
\gamma _{\pm }^{(0)}(j)=-4S_{1}(j\mp 2),~~~~\gamma
_{0}^{(0)}(j)=-4S_{1}(j),~~~~\widetilde{\gamma}_{\pm }^{(0)}(j)=-4S_{1}(j\mp
1)\,.
\]

To illustrate the procedure we consider the polarized and unpolarized cases
separately.\newline

{\it Polarized case.} From Eqs. (\ref{3.34}), (\ref{3.4}) and (\ref{fu.2})
we have
\begin{eqnarray}
&&\hspace*{-1cm} \widetilde \gamma^{(0)}_{gg}(j) + \frac{1}{2} \widetilde
\gamma^{(0)}_{qg}(j) ~=~ \widetilde \gamma^{(0)}_{qq}(j) + 2 \widetilde
\gamma^{(0)}_{gq}(j) ~=~ \widetilde \gamma^{(0)}_{+}(j)\equiv -4S_1(j-1)
\label{fu.3} \\
&&\hspace*{-1cm}  \nonumber \\
&&\hspace*{-1cm} 
\widetilde \gamma^{(0)}_{gg}(j) - \frac{j+2}{2(j-1)} \widetilde
\gamma^{(0)}_{qg}(j) ~=~ \widetilde \gamma^{(0)}_{qq}(j) - \frac{2(j-1)}{j+2}
\widetilde \gamma^{(0)}_{gq}(j) ~=~ 
\widetilde \gamma^{(0)}_{-}(j) \equiv -4S_1(j+1)
\label{fu.4}
\end{eqnarray}

It is obvious, that Eqs.(\ref{3.31})-(\ref{3.4}) and
(\ref{fu.3})-(\ref{fu.4}%
) correspond to a diagonalization of the anomalous dimension matrix. So we
can rewrite Eqs. (\ref{fu.3})-(\ref{fu.4}) as
\begin{eqnarray}
{\hat{\widetilde V}}^{-1} \left(
\begin{array}{cc}
\widetilde \gamma^{(0)}_{gg}(j) & \widetilde \gamma^{(0)}_{qg}(j) \\
\widetilde \gamma^{(0)}_{gq}(j) & \widetilde \gamma^{(0)}_{qq}(j)
\end{array}
\right) \hat{\widetilde V} ~=~ \left(
\begin{array}{cc}
\widetilde \gamma^{(0)}_{+}(j) & 0 \\
0 & \widetilde \gamma^{(0)}_{-}(j)
\end{array}
\right) ~~~~~ \left({\hat{\widetilde V}}^{-1} \hat{\widetilde V}=1 \right),
\label{fu.5}
\end{eqnarray}
where
\begin{eqnarray}
\hat{\widetilde V} ~=~ \left(
\begin{array}{rr}
\widetilde v_g & -2\frac{j-1}{j+2}\widetilde v_q \\
\frac{1}{2} \widetilde v_g & \widetilde v_q
\end{array}
\right),~~ {\hat{\widetilde V}}^{-1} ~=~ \frac{j+2}{2j+1} \left(
\begin{array}{rr}
\widetilde v_g^{-1} & 2\frac{j-1}{j+2}\widetilde v_g^{-1} \\
-\frac{1}{2} \widetilde v_q^{-1} & \widetilde v_q^{-1}
\end{array}
\right)  \label{fu.6}
\end{eqnarray}
with arbitrary values of $\widetilde v_g $ and $\widetilde v_q $.

Thus, Eq.(\ref{fu.5}) leads to the following representation of the anomalous
dimension matrix in the polarized case
\begin{eqnarray}
&&\hspace*{-1cm} \hat {\widetilde \gamma}^{(0)}(j) \equiv \left(
\begin{array}{cc}
\widetilde \gamma^{(0)}_{gg}(j) & \widetilde \gamma^{(0)}_{qg}(j) \\
\widetilde \gamma^{(0)}_{gq}(j) & \widetilde \gamma^{(0)}_{qq}(j)
\end{array}
\right) ~=~ \hat{\widetilde V} \left(
\begin{array}{cc}
\widetilde \gamma^{(0)}_{+}(j) & 0 \\
0 & \widetilde \gamma^{(0)}_{-}(j)
\end{array}
\right) {\hat{\widetilde V}}^{-1}  \nonumber \\
&&\hspace*{-1cm}  \nonumber \\
&&\hspace*{-1cm} = \frac{1}{2(2j+1)} \left[ \widetilde \gamma^{(0)}_{+}(j)
\left(
\begin{array}{cc}
2(j+2) & 4(j-1) \\
j+2 & 2(j-1)
\end{array}
\right) ~+~ \widetilde \gamma^{(0)}_{-}(j) \left(
\begin{array}{cc}
2(j-1) & -4(j-1) \\
-(j+2) & 2(j+2)
\end{array}
\right) \right]  \label{fu.7}
\end{eqnarray}
which does not depend on $\widetilde v_g $ and $\widetilde v_q $.\newline


{\it Unpolarized case.} From Eqs. (\ref{3.34}), (\ref{3.4}) and (\ref{fu.1})
we have
\begin{eqnarray}
&&\hspace*{-1cm} \gamma^{(0)}_{gg}(j) + \gamma^{(0)}_{qg}(j) +
\gamma^{(0)}_{\varphi g}(j) ~=~ \gamma^{(0)}_{qq}(j) + \gamma^{(0)}_{gq}(j)
+ \gamma^{(0)}_{\varphi q}(j)  \nonumber \\
&&\hspace*{-1cm} ~=~ \gamma^{(0)}_{\varphi \varphi}(j) + \gamma^{(0)}_{g
\varphi}(j) + \gamma^{(0)}_{q\varphi}(j) ~=~ \gamma^{(0)}_{+}(j) \equiv
-4S_1(j-2) ,  \label{fu.10} \\
&&\hspace*{-1cm}  \nonumber \\
&&\hspace*{-1cm} \gamma^{(0)}_{gg}(j) - \frac{1}{2(j-1)}
\gamma^{(0)}_{qg}(j) - \frac{j+1}{3(j-1)} \gamma^{(0)}_{\varphi g}(j)
\nonumber \\
&&\hspace*{-1cm} ~=~ \gamma^{(0)}_{qq}(j) - 2(j-1) \gamma^{(0)}_{gq}(j) +
\frac{2(j+1)}{3} \gamma^{(0)}_{\varphi q}(j)  \nonumber \\
&&\hspace*{-1cm} ~=~ \gamma^{(0)}_{\varphi \varphi}(j) - \frac{3(j-1)}{j+1}
\gamma^{(0)}_{g\varphi}(j) + \frac{3}{2(j+1)} \gamma ^{(0)}_{q\varphi }(j)
~=~ \gamma^{(0)}_{0}(j) \equiv -4S_1(j),  \label{fu.11} \\
&&\hspace*{-1cm}  \nonumber \\
&&\hspace*{-1cm} \gamma^{(0)}_{gg}(j) - \frac{j+2}{(j-1)}
\gamma^{(0)}_{qg}(j) + \frac{(j+1)(j+2)}{j(j-1)} \gamma^{(0)}_{\varphi g}(j)
\nonumber \\
&&\hspace*{-1cm} ~=~ \gamma^{(0)}_{qq}(j) - \frac{j-1}{j+2}
\gamma^{(0)}_{gq}(j) - \frac{j+1}{j} \gamma^{(0)}_{\varphi q}(j)  \nonumber
\\
&&\hspace*{-1cm} ~=~ \gamma^{(0)}_{\varphi \varphi}(j) + \frac{j(j-1)}{%
(j+1)(j+2)} \gamma^{(0)}_{g\varphi}(j) - \frac{j}{j+1} \gamma^{(0)}_{q%
\varphi }(j) ~=~ \gamma^{(0)}_{-}(j) \equiv -4S_1(j+2).  \label{fu.12}
\end{eqnarray}

The formulae (\ref{fu.10})-(\ref{fu.12}) are equivalent to the matrix
equation
\begin{eqnarray}
{\hat{V}}^{-1} \left(
\begin{array}{ccc}
\gamma^{(0)}_{gg}(j) & \gamma^{(0)}_{qg}(j) & \gamma^{(0)}_{\varphi g}(j) \\
\gamma^{(0)}_{gq}(j) & \gamma^{(0)}_{qq}(j) & \gamma^{(0)}_{\varphi q}(j) \\
\gamma^{(0)}_{g\varphi }(j) & \gamma^{(0)}_{q\varphi }(j) &
\gamma^{(0)}_{\varphi \varphi}(j)
\end{array}
\right) \hat{V} ~=~ \left(
\begin{array}{ccc}
\gamma^{(0)}_{+}(j) & 0 & 0 \\
0 & \gamma^{(0)}_{0}(j) & 0 \\
0 & 0 & \gamma^{(0)}_{-}(j)
\end{array}
\right) ~ \left({\hat{ V}}^{-1} \hat{ V}=1 \right),  \label{fu.13}
\end{eqnarray}
where
\begin{eqnarray}
\hat{V} &=& \left(
\begin{array}{ccc}
v_g & -2(j-1)v_q & \frac{j(j-1)}{(j+1)(j+2)} v_{\varphi} \\
v_g & v_q & -\frac{j}{j+1} v_{\varphi} \\
v_g & \frac{2}{3}(j+1) v_q & v_{\varphi}
\end{array}
\right),~~  \label{fu.14} \\
& &  \nonumber \\
{\hat{V}}^{-1} &=& \frac{(j+1)(j+2)}{2(4j^2-1)(2j+3)} \left(
\begin{array}{ccc}
(2j+3)v_g^{-1} & 4\frac{j-1}{j+2}(2j+3)v_g^{-1} &
3\frac{j(j-1)}{(j+1)(j+2)}%
(2j+3)v_g^{-1} \\
-3\frac{2j+1}{j+1} v_q^{-1} & 6\frac{2j+1}{(j+1)(j+2)} v_q^{-1} & \frac{%
j(2j+1)}{(j+1)(j+2)}v_q^{-1} \\
(2j-1)v_{\varphi}^{-1} & 4(2j-1)v_{\varphi}^{-1} & 3(2j-1)v_{\varphi}^{-1}
\end{array}
\right)  \nonumber
\end{eqnarray}
with arbitrary values of $v_g $, $v_q $ and $v_{\varphi} $.

Analogously to the polarized case one can write the expression for the
anomalous dimension matrix in the form
\begin{eqnarray}
&&\hspace*{-1cm} \hat {\gamma}^{(0)}(j) \equiv \left(
\begin{array}{ccc}
\gamma^{(0)}_{gg}(j) & \gamma^{(0)}_{qg}(j) & \gamma^{(0)}_{\varphi g}(j) \\
\gamma^{(0)}_{gq}(j) & \gamma^{(0)}_{qq}(j) & \gamma^{(0)}_{\varphi q}(j) \\
\gamma^{(0)}_{g\varphi }(j) & \gamma^{(0)}_{q\varphi }(j) &
\gamma^{(0)}_{\varphi \varphi}(j)
\end{array}
\right) ~=~ \hat{V} \left(
\begin{array}{ccc}
\gamma^{(0)}_{+}(j) & 0 & 0 \\
0 & \gamma^{(0)}_{0}(j) & 0 \\
0 & 0 & \widetilde \gamma^{(0)}_{-}(j)
\end{array}
\right) {\hat{V}}^{-1}  \nonumber \\
&&\hspace*{-1cm}  \nonumber \\
&&\hspace*{-1cm} = \frac{1}{2(4j^2-1)(2j+3)} \Biggl[ (2j+3)
\,\gamma^{(0)}_{+}(j) \cdot \left(
\begin{array}{ccc}
(j+1)(j+2) & 4(j^2-1) & 3j(j-1) \\
(j+1)(j+2) & 4(j^2-1) & 3j(j-1) \\
(j+1)(j+2) & 4(j^2-1) & 3j(j-1)
\end{array}
\right)  \nonumber \\
& &  \nonumber \\
&&\hspace*{-1cm} ~+~ (2j+1) \,\gamma^{(0)}_{0}(j) \cdot \left(
\begin{array}{ccc}
6(j-1)(j+2) & -12(j-1) & -6j(j-1) \\
-3(j+2) & 6 & 3j \\
-2(j+1)(j+2) & 4(j+1) & 2j(j+1)
\end{array}
\right)  \nonumber \\
&&\hspace*{-1cm}  \nonumber \\
&&\hspace*{-1cm} ~+~ (2j-1) \,\gamma^{(0)}_{-}(j) \cdot \left(
\begin{array}{ccc}
j(j-1) & 4j(j-1) & 3j(j-1) \\
-j(j+2) & -4j(j+2) & 3j(j+2) \\
(j+1)(j+2) & 4(j+1)(j+2) & 3(j+1)(j+2)
\end{array}
\right) \Biggr]  \label{fu.15}
\end{eqnarray}
which does not depend on $v_g $, $v_q $ and $v_{\varphi} $.\newline

\section{Appendix D}

\label{App:D}
\def\theequation{D\arabic{equation}}
\setcounter{equation}0

Here we obtain the expression (\ref{BSam1}) for the $t$-channel partial wave
$f_{\omega }(\kappa )$ of the process $e^+e^{-}\rightarrow \mu ^{+}\mu ^{-}$ 
in the double-logarithmic approximation. 
Note that our results coincide with those obtained in the 
original paper \cite{FGGL}.

The Bethe-Salpeter equation (\ref{BSam})

\[
A(s,k_{\perp }^{2})=1+\lambda \int_{m^{2}}^{s}\frac{dk_{\perp }^{\prime 2}}{%
k_{\perp }^{\prime 2}}\int_{k_{\perp }^{\prime 2}}^{\min (s,sk_{\perp
}^{\prime 2}/k_{\perp }^{2})}\frac{ds^{\prime }}{s^{\prime }}\,A(s^{\prime
},k_{\perp }^{\prime 2})\,\,\,\,
~~\left(\lambda =\frac{\alpha _{em}}{2\pi }\right)
\]
can be transformed after going in the $\omega $-representation
\[
A(s,k_{\perp }^{2})=\int_{\sigma -i\infty }^{\sigma +i\infty }\frac{d\omega
}{2\pi i\,\omega }\,\left( \frac{s}{k_{\perp }^{2}}\right) ^{\omega
}f_{\omega }(\kappa )\,\,\,
~~\left(\kappa =\ln \frac{k_{\perp }^{2}}{m^{2}}\right)
\]
to the form
\[
f_{\omega }(\kappa )=1+\frac{\lambda }{\omega }\left( \int_{\kappa
}^{\Lambda }d\kappa ^{\prime }\,e^{-\omega \,(\kappa ^{\prime }-\kappa
)}\,f_{\omega }(\kappa ^{\prime })+\int_{0}^{\kappa }d\kappa ^{\prime
}\,f_{\omega }(\kappa ^{\prime })\right) \,.
\]

We introduced here the ultraviolet cut-off $\Lambda $ in $\ln k^2_{\perp }$.
It will be pushed to infinity in the end of calculations. Note, that in QCD
such cut-off could be interpreted as an effect of the interaction vanishing at
large $k^2$ due to the asymptotic freedom. The
differentiation of the integral equation gives

\[
e^{-\omega \kappa }\frac{d}{d\kappa }f_{\omega }(\kappa )=\lambda
\int_{\kappa }^{\Lambda }d\kappa ^{\prime }\,e^{-\omega \kappa ^{\prime
}}\,f_{\omega }(\kappa ^{\prime })\,.
\]
The second differentiation leads to the differential equation

\[
\left( e^{\omega \kappa }\frac{d}{d\kappa }e^{-\omega \kappa }\frac{d}{%
d\kappa }+\lambda \right) \,f_{\omega }(\kappa )=0\,.
\]

We can find its solution in terms of two arbitrary constants $a_{\omega
}^{\pm }$
\[
f_{\omega }(\kappa )=a_{\omega }^{-}\,e^{\gamma _{\omega }^{-}\,\kappa
}+a_{\omega }^{+}\,e^{\gamma _{\omega }^{+}\,\kappa }\,,
\]
where
\bea
\gamma _{\omega }^{\pm }~=~
\frac{\omega }{2}\,\, \left( 1\pm \sqrt{1-4\frac{\lambda
}{\omega ^{2}}}\right) \,.
\label{BSam3}
\eea

>From the above expression for $e^{-\omega \kappa }\frac{d}{d\kappa }%
f_{\omega }(\kappa )$ we obtain, that
\[
a_{\omega }^{-}\,\gamma _{\omega }^{-}e^{\gamma _{\omega }^{-}\,\Lambda
}+a_{\omega }^{+}\,\gamma _{\omega }^{+}e^{\gamma _{\omega }^{+}\Lambda
\,}\,=0,
\]
and from the integral equation
\[
1\, = \, \frac{\lambda }{\omega } \,
\left( \frac{a_{\omega }^{-}}{\gamma _{\omega
}^{-}\,}+\frac{a_{\omega }^{+}}{\gamma _{\omega }^{+}\,}\,\right) .
\]

Therefore
\[
a_{\omega }^{+}=\frac{\omega }{\lambda } \,\,
e^{\Lambda \gamma _{\omega }^{-}} \,
{\left[\frac{1}{\gamma _{\omega }^{+}}
e^{\Lambda \gamma _{\omega }^{-}\,}-
\frac{\gamma _{\omega }^{+}}{\gamma _{\omega }^{-2}}
e^{\Lambda \gamma _{\omega }^{+}}\right]}^{-1}\,,\,\,~
a_{\omega }^{-}=\frac{\omega }{\lambda } \,\,
e^{\Lambda \gamma _{\omega }^{+}} \,
{\left[
\frac{1}{\gamma _{\omega }^{-}}%
e^{\Lambda \gamma _{\omega }^{+}\,}-
\frac{\gamma _{\omega }^{-}}{\gamma_{\omega }^{+2}}
e^{\Lambda \gamma _{\omega }^{-}}\right]}^{-1}\,.
\]
In particular, for $\Lambda \rightarrow \infty $ we obtain
\[
f_{\omega }(\kappa )=f_{\omega }\,e^{\gamma _{\omega }^{-}\,\kappa },\,\,
\]
where the $t$-channel partial wave on the mass shell

\bea
f_{\omega }~=~\frac{\omega }{\lambda }\,
\gamma _{\omega }^{-} \, = \, 
\frac{\omega ^{2}}{2\lambda }\,
\left( 1-\sqrt{1-4\frac{\lambda }{\omega ^{2}}}\right)
\label{BSam4}
\eea
satisfies the infrared evolution equation
\bea
f_{\omega }=1+\frac{\lambda }{\omega ^{2}}\,f_{\omega }^{2}\,.
\label{BSam5}
\eea
In this case the partial wave can be written also as follows
\bea
f_{\omega}(\kappa )=
\int _{\frac{\omega}{2}-i\infty}^{\frac{\omega}{2}+i\infty}\frac{d \gamma}{2\pi i}\,
\left(\frac{k^2}{m^2}\right)^{\gamma}
\frac{(\omega -2\gamma )\,\gamma \,\omega ^2 \,\lambda ^{-2}}{\omega -
\frac{\alpha _{em}}{2\pi}\left[\frac{1}{\omega -\gamma}+\frac{1}{\gamma}\right]}
\label{BSam6}
\eea

Note, that providing that $\Lambda =\infty $ from the beginning the
condition $a_{\omega }^{+}=0$ follows from the requirement, that the 
$t$-channel partial wave $f_{\omega }(\kappa )$ does not grow at $\omega
\rightarrow \infty $ (in an opposite case we can not reduce the initial
Bethe-Salpeter equation for $A(s,k_{\perp }^{2})$ to the equation for $%
f_{\omega }(\kappa )$ because the lower limit of integration $s^{\prime
}=k_{\perp }^{\prime 2}$ is not zero).

Thus, we derived the
correct expression (\ref{BSam1})
for $f_\omega (\gamma)$. The '$-$' component of the value (\ref{BSam3}) coincides with
one (\ref{BSam2}) in the section 2.




\end{document}